\documentclass[twocolumn,showpacs,preprintnumbers,superscriptaddress,amsmath,amssymb]{revtex4-1}

\usepackage[colorlinks=true,bookmarks=false,citecolor=blue,urlcolor=blue]{hyperref} 

\usepackage{amsmath,amssymb}
\usepackage{graphicx}
\usepackage{verbatim}
\usepackage{enumitem}
\usepackage{color}
\usepackage[english]{babel}
\usepackage{bm}
\usepackage{natbib}
\usepackage[symbol]{footmisc}
\usepackage{textcomp}
\newcommand{\red}{\textcolor{black}}

\usepackage{epstopdf}
\usepackage{setspace}
\usepackage{comment}

\begin{document}

\title{Halide-Perovskite Resonant Nanophotonics}

\author{Sergey Makarov}
\affiliation{ITMO University, St.~Petersburg 197101, Russia}
\author{Aleksandra Furasova}
\affiliation{ITMO University, St.~Petersburg 197101, Russia}
\author{Ekaterina Tiguntseva}
\affiliation{ITMO University, St.~Petersburg 197101, Russia}
\author{Andreas Hemmetter}
\affiliation{ITMO University, St.~Petersburg 197101, Russia}
\author{Alexander Berestennikov}
\affiliation{ITMO University, St.~Petersburg 197101, Russia}
\author{Anatoly Pushkarev}
\affiliation{ITMO University, St.~Petersburg 197101, Russia}
\author{Anvar Zakhidov}
\affiliation{ITMO University, St.~Petersburg 197101, Russia}
\affiliation{University of Texas at Dallas, Richardson TX 75080, USA}
\author{Yuri Kivshar}
\affiliation{ITMO University, St.~Petersburg 197101, Russia}
\affiliation{Nonlinear Physics Centre, Australian National University, Canberra ACT 2601, Australia}

\begin{abstract}
Halide perovskites have emerged recently as promising materials for many applications in photovoltaics and optoelectronics. Recent studies of their optical properties suggest many novel opportunities for a design of advanced nanophotonic devices due to low-cost fabrication, high values of the refractive index,  existence of excitons at room temperatures, broadband bandgap tunability, high optical gain and nonlinear response, as well as simplicity of their integration with other types of structures. This paper provides an overview of the recent progress in the study of optical effects originating from nanostructured perovskites, including their potential applications.
\end{abstract}

\maketitle

\section{Introduction}

Optically resonant nanostructures provide a bridge between optics and nanoscale sciences, allowing for shrinking light confinement down to the nanoscale via excitation of highly localized optical modes. This enables to scale down a number of important optical devices such as waveguides, lasers, photodetectors, sensors, etc. Historically, metal nanostructures paved the way to novel nanoscale optical phenomena and applications related to the effective light management in the deeply subwavelength regime~\cite{moskovits1985surface, ozbay2006plasmonics, atwater2010plasmonics, zheludev2012metamaterials}. The materials used include gold, silver, and copper~\cite{murray2007plasmonic}, as well as various metal alloys and doped oxides~\cite{west2010searching}. Furthermore, to overcome optical losses and bring novel functionality, dielectric resonant nanostructures have been introduced and extensively studied over the last decade~\cite{kuznetsov2016optically, eaton2016semiconductor}, where conventional inorganic materials such as silicon, gallium arsenide, gallium nitride, zinc oxide, etc. were employed due to their high values of the refractive index and well-developed methods of fabrication. However, most of these materials face some limitations related to difficulties with spectral tunability, a lack of excitons at room temperatures, expensive fabrication processes, and a low quantum yield.

\begin{figure}[h!]
 \centering
\includegraphics[width=0.85\linewidth]{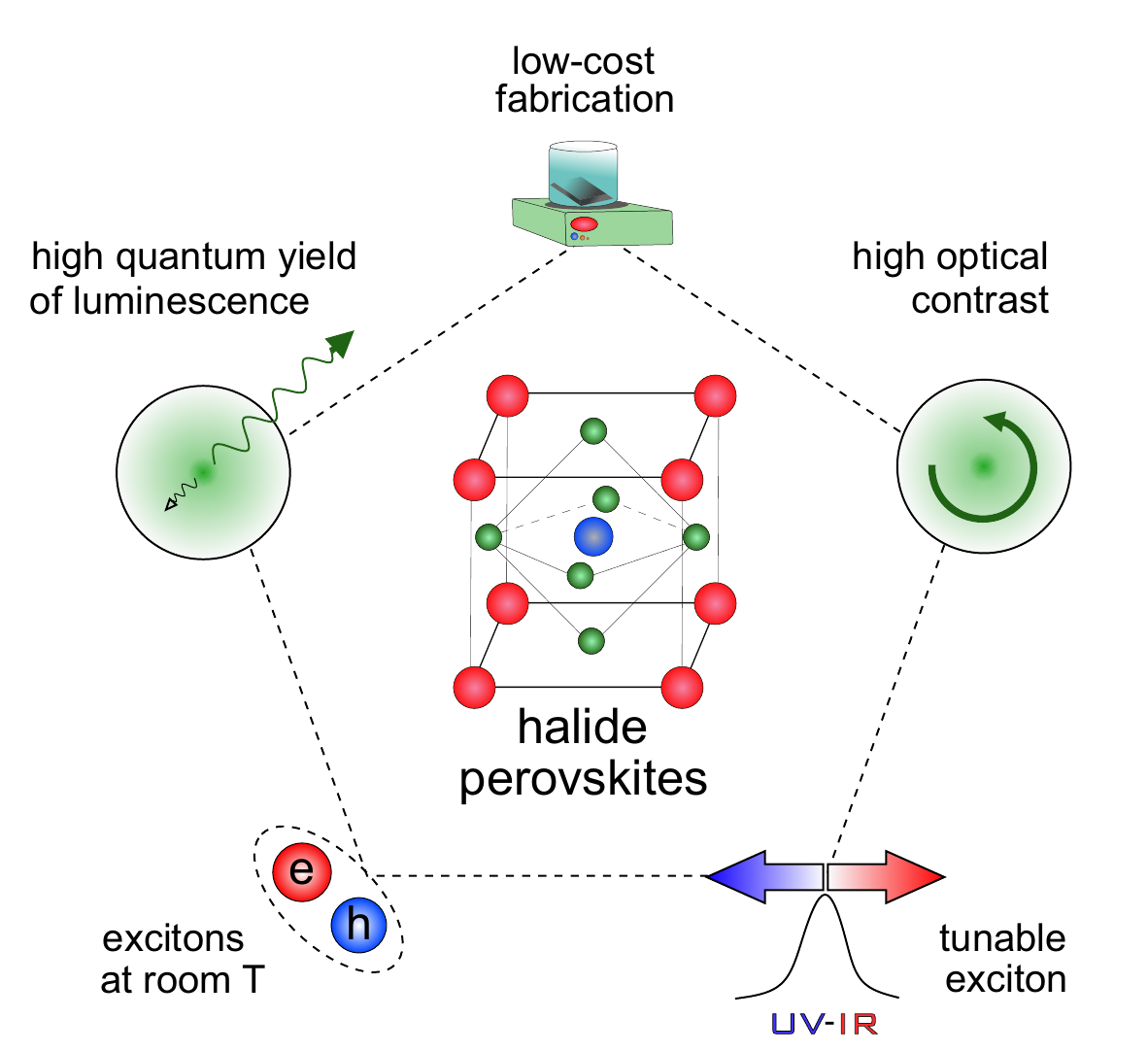}
\caption{\textbf{Schematic illustration of major properties} of halide perovskites  emerged recently as promising novel materials for many applications in optoelectronics and photonics.
}
\label{material}
\end{figure}

\begin{figure*}
 \centering
\includegraphics[width=0.85\linewidth]{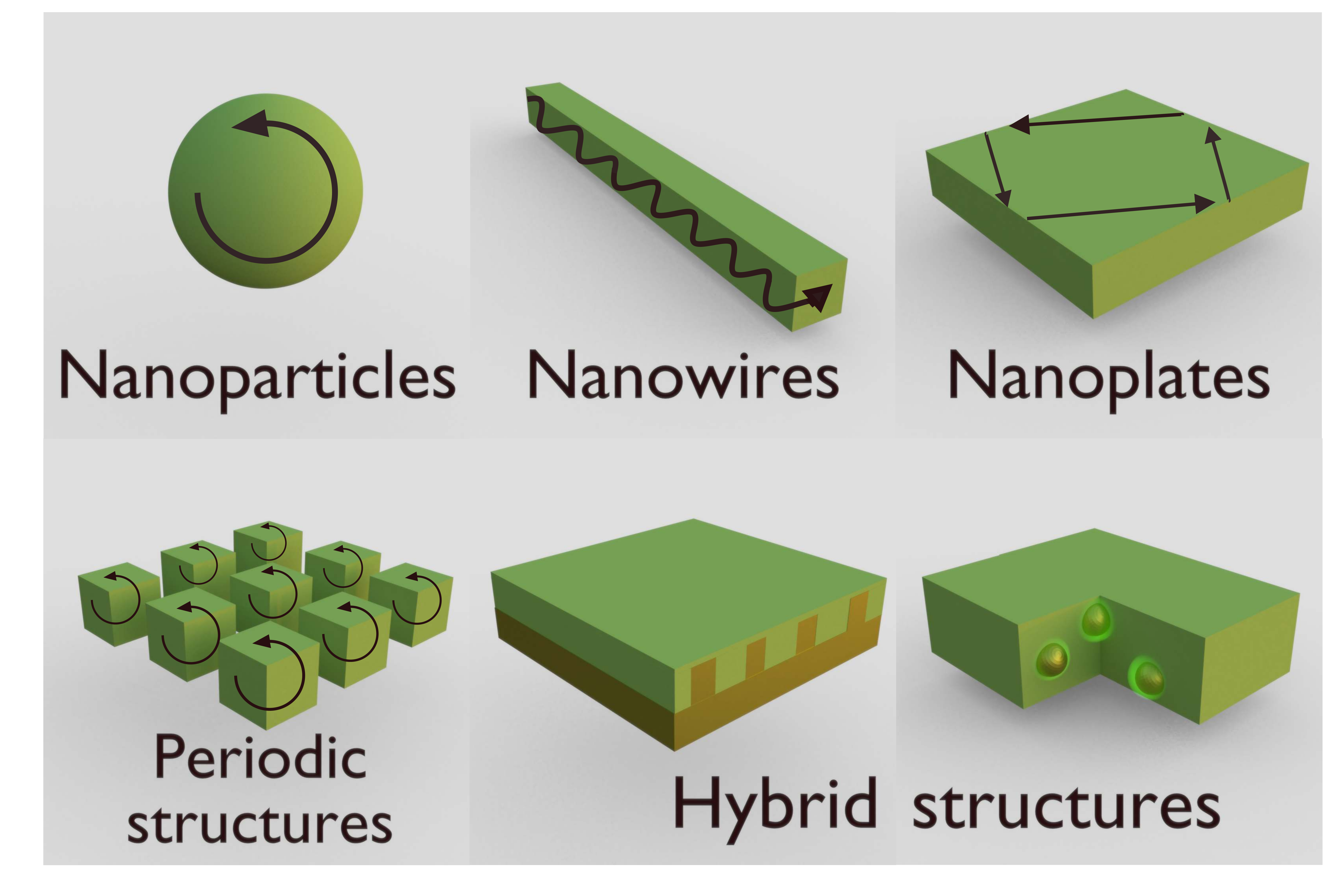}
\caption{\textbf{Basic designs of resonant nanostructures made of halide perovskites.} Single nanoparticles supporting the Mie resonances (Section 3.1). Nanowires supporting the Fabry-Perot resonances and waveguide modes (Section 3.2). Nanoplates supporting whispering-gallery modes (Section 3.2). Two-dimensioanl arrays of resonant nanoparticles creating either photonic crystals or metasurfaces (Section 3.3). Layer of halide perovskite can be deposited on resonant substrates (Sections 3.3 and 3.4). Resonant non-perovskite nanoparticles  combined with a perovskite layer or incorporated into a perovskite slab, respectively (Section 3.4 and Section 4).}
\label{designs}
\end{figure*}

On the other hand, a new class of materials, the so-called {\it halide perovskites},
has emerged recently and attracted a lot of attention not only for photovoltaics~\cite{green2014emergence}, but also for photonic sources~\cite{sutherland2016perovskite}. The main reason for such an interest is the outstanding electronic and photonic properties of halide perovskites, along with low-cost of their fabrication, extremely broadband spectral tunability, and other properties and features, as shown schematically in Fig.~\ref{material}. Additionally, Fig.~\ref{designs} summarizes differenmt types of resonant nanostructures made of halide perovskites (such as nanoparticles, nanowires, nanoplates, photonic crystals, and metasurfaces) as well as created by integration of perovskites with other non-perovskite nanophotonic structures, which implement all mentioned advantages of the perovskites. 

In this review paper, we introduce the recently emerged field of nanophotonics based on halide perovskites, describe main (linear and nonlinear) optical properties of this class of materials, explain why fundamental discoveries in the halide perovskite nanophotonics are important for photonics devices, and offer our perspective on the future directions in this actively developing research field. 

\section{Material properties}

\subsection{Structure and chemistry}

The first sample of a mineral with a  perovskite structure, namely, CaTiO$_3$ was found by Gustav Rose in the Ural mountains in 1839~\cite{rose39}. Its generalization is a compound with the stoichiometric formula ABX$_3$, where $A$ and $B$ are large and small cations, respectively, and $X$ is an anion, and they all crystallize in the same structure as CaTiO$_3$. In the ideal case of a cation $B$ much smaller than cation $A$, perovskites assemble in {\it a cubic lattice} with the coordination numbers 12 for $A$, 6 for $B$, and 8 for $X$. Due to strict constraints in the ionic radii, only few perovskites have this ideal cubic structure. The Goldschmidt tolerance factor $t = \frac{r_A + r_X}{\sqrt{2}(r_B + r_X)},$ where $r$ stands for the radius of the corresponding ion, is a measure for the degree to which a given compound crystallizes in the ideal perovskite structure. If $t<0.89$ or $t>1.02$, the structure has a reduced symmetry in the form of an orthorhombic or tetragonal structure with lower coordination numbers~\cite{goldschmidt26}. The perovskite structure can be adopted by a wide range of compounds in the above mentioned stoichiometric composition ABX$_3$.

For novel photonic applications, very important compounds are  {\it halide perovskites}, which contain an organic or inorganic compound as $A$ cation (e.g. methylammonium (MA$^+$), formamidimium (FA$^+$) or Cs$^+$), a metallic $B$ cation (e.g. Pb$^{2+}$ or Sn$^{2+}$) and a variable composition of halide anions $X$ (Cl$^-$, Br$^-$, I$^-$). Owing to their good tolerance factor (\textit{t}$\approx$1), and widespread use in photonics, photovoltaics, and optoelectronics, this review discusses mainly the nanostructures based on inorganic and hybrid lead halide perovskites, i.e (MA,FA,Cs)Pb(Cl,Br,I)$_3$.

\begin{figure*}
 \centering
\includegraphics[width=1.0\linewidth]{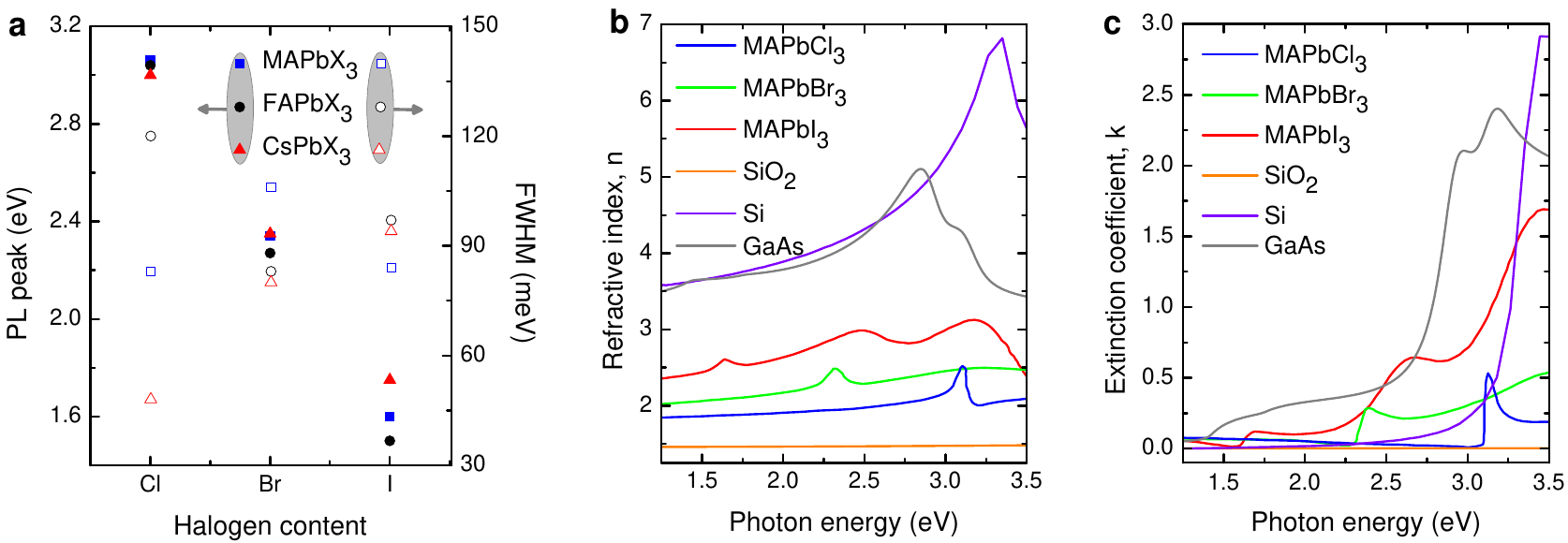}
\caption{\footnotesize{\textbf{Optical properties of halide perovskites.}} \red{(a) Photoluminescence peak energy and FWHM for pure chloride, bromide, and iodide perovskite thin films in the cubic phase.} \red{(b)} Refractive indices $n$ and \red{(c)} extinction coefficients $k$ of three 
commonly used perovskite composites: MAPbCl$_3$, MAPbBr$_3$, and MAPbI$_3$, shown together with the corresponding results for SiO$_2$, Si, and GaAs, for comparison. The data for the perovskites are obtained by Leguy {\it et al.} through single crystal ellipsometry and from Refs.~\cite{rodriguez2016self,schinke2015uncertainty,ozaki1995spectroscopic} for SiO$_2$, Si, and GaAs. These data are collected from Refs.~\cite{wang2016visible, kim2017high, ko201515, imran2018benzoyl, slimi2017synthesis, yuan2015inverted, kondo2005room, sebastian2015excitonic, wang2017stabilizing}.}
\label{Optics}
\end{figure*}

\subsection{Linear optics with perovskites}

Optical characteristics of perovskites can be varied by changing the anion X$^-$, because of the dependence of the valence and conduction bands of APbX$_3$ materials on lead-halogen ionic binding strength decreasing in the sequence Pb-Cl $<$ Pb-Br $<$ Pb-I in accord with halogens electronegativity. The bandgaps for MAPbX$_3$ where a single type of halogen atom occupies the X site are very different: 3.09 (X = Cl), 2.32 (X = Br) and 1.7~eV (X = I)~\cite{li2016halide}, but quite similar to those for CsPbX$_3$: 3.0 (X = Cl)~\cite{heidrich1981electronic}, 2.39 (X = Br)~\cite{ng2018tunable} and 1.73~eV (X = I)~\cite{ahmad2017inorganic}. For the MAPbX$_3$ this corresponds to \red{blue wing of the emission peak (Fig.~\ref{Optics}a)} and onset of absorption near 400, 530 and 730~nm \red{(Fig.~\ref{Optics}c)}, respectively. Moreover, the emission and absorption peaks can be tuned finer by including stoichiometric mixtures of two halides. Such mixtures, for example, MAPb(I$_{1-x}$Br$_x$), where $0 \leq x \leq 1$, was shown to be able to tune the bandgap, and, therefore, emission peak and absorption onset, to an arbitrary value between the limit cases~\cite{stranks15,jeon15,manser16}. Since the energy states contributed by the large cation were calculated to be far from the band edges, A$^+$ \red{has weak} impact on the electronic structure of perovskites and plays a role of a crystal lattice stabilizing unit~\cite{yin2015halide}.

The important feature of the emission of halide perovskites is their excitonic nature. A number of studies reported contradicting results on the Wannier-Mott exciton binding energies \red{around 41~meV~\cite{yamada2018near} for MAPbCl$_3$; 15-40~meV~\cite{tanaka2003comparative,kunugita2015excitonic,yang2015comparison, tilchin2016hydrogen} for MAPbBr$_3$; 
and 5-15~meV~\cite{miyata2015direct,yang2016observation} for MAPbI$_3$. The uncertainties} of the values can be attributed to a difference in the studied samples and their morphology.

\red{These values can be slightly increased by the contraction of the crystal lattice via introduction of Cs$^+$ cation, instead of organic cation. The binding energies of excitons for CsPbCl$_3$ and CsPbBr$_3$ perovskites are 72~meV and 38~meV, respectively~\cite{wang2016high}. Therefore, excitons in most of perovskite objects} can survive at room temperatures and yield narrow and efficient photoluminescence in the wide spectral range [see Fig.~\ref{Optics}(a)] \red{which is} useful for many applications in photonics.

\begin{table}[h!]
\centering
\caption{Excitonic parameters of hybrid halide perovskites}
\label{par}
\begin{tabular}{lllllll}
\hline
Compound    & $\omega_{exc}$(eV) & $\mu_{exc}$ (m$_e$) &  $\omega_p$(eV) & $\gamma$(eV) \\ \hline
MAPbI$_3$         & 1.55                 & 0.104                          & 0.35       & 0.08     \\
MAPbBr$_3$         & 2.28                 & 0.113                          & 0.45       & 0.06     \\
MAPbCl$_3$         & 3.05                 & 0.136                         & 0.85       & 0.05    
\end{tabular}
\end{table}

\red{The} dependence of the dielectric permittivity $\varepsilon$ of materials with a strong excitonic contribution vs. frequency ($\omega$) and the wave vector (\textit{k})~\cite{agranovich}, taking into account the spatial dispersion is given by the relation:
\begin{equation}
\varepsilon (\omega, k) = \varepsilon_0 + \frac{\omega_p^2}{\omega_{exc}^2 - \omega^2 + Dk^2 - i\gamma\omega}\ ,
\end{equation}
where $\omega_{exc}$ is the frequency of excitonic transition, $\omega_{p}$ is the strength of a dipole oscillator, $\gamma$ is the damping factor, $\varepsilon_0$ is the background dielectric constant, and $\mathrm{D}k^2 = \hbar^2k^2\omega_{exc}/\mu_{exc}$ is a term related to a nonlocal response, \red{being comparable with extensively studied analogous nonlocality effects from free electrons in metallic nanoparticles~\cite{krasavin2018free}.} 
\begin{table*}
\centering
\caption{Lasing characteristics of hybrid halide perovskites}
\label{gain}
\begin{tabular}{|l|l|l|l|l|}
\hline
Compound   & Lasing wavelength & Gain (cm$^{-1}$) & \begin{tabular}[c]{@{}l@{}}ASE threshold at pulsed pump\\ (J/cm$^2$) with given wavelength\end{tabular} & Ref.                                        \\ \hline
MAPbI$_3$  & $\approx$760--790~nm         & 250             & 10$^{-5}$--10$^{-4}$  @ $\lambda_{p}$=400~nm                                                                             & \cite{xing2014low}       \\ \hline
MAPbBr$_3$ & $\approx$530--550~nm         & 300             & 10$^{-5}$--10$^{-4}$  @ $\lambda_{p}$=355~nm                                                                            & \cite{lafalce2016enhanced} \\ \hline
MAPbCl$_3$ & $\approx$400--410~nm         & 110             & 10$^{-2}$ @ $\lambda_{p}$=2100~nm                                                                          & \cite{yang2018lasing}      \\ \hline
\end{tabular}
\end{table*}

\red{In general, halide perovskites have a mid-range refractive index in the visible range (Fig.~\ref{Optics}b) and a sharp increase of losses near the exciton state [see Fig.~\ref{Optics}(c)]. Unlike their bandgap, the refractive index depends largerly on the crystal morphology. MAPbI$_3$ as a single crystal, and it has the refractive index of 2.45 at 800~nm, while the same material at the same wavelength shaped as a thin film only has $n = 1.95$. To the best of our knowledge, there are no reliable experimental data on the values of the refractive index of CsPbX$_3$. Importantly, the real part of the refractive index of perovskites is significantly larger than that for SiO$_2$ or most of polymers, making perovskites a good material for resonant nanostructures, due to high enough optical contrast. On the other hand, it is much lower than those for Si and GaAs, providing high optical contrast with these materials in advanced hybrid structures.}

Another property of perovskites is high absorption coefficient at the energies higher than that of the exciton state [see Fig.~\ref{Optics}(c)], which is desirable for photovoltaic and lasing applications. Absorption coefficients up to $2\times10^5$~cm$^{-1}$ in the visible fraequency range have been reported~\cite{mitzi07}.

\subsection{Lasing properties} 

Perovskites possess a high optical gain and wavelength tunability in the range of 400--1000~nm~\cite{zhang2017advances}. The visible range is covered by APbX$_3$ perovskites, whereas in the near-IR range lasing is achieved with CsSnI$_3$ perovskite material~\cite{xing2016solution}. One of the highest values of the perovskite gain is 3200~cm$^{-1}$ reported in Ref.~\cite{sutherland2015perovskite}, and it depends strongly on film quality affecting excitons binding energies and mobilities. The \red{typical} lasing characteristics of hybrid halide perovskites are summarized in Table~\ref{gain}. These gain values obtained from halide perovskites compare favourably with those for colloidal semiconductor QDs~\cite{dang2012red} and conjugated polymer thin films~\cite{lampert2012controlling} at comparable excitation intensities. In Sec.~3, we discuss lasing characteristics of various resonant nanostructures.

\subsection{Nonlinear optics with perovskites}

\textbf{Nonlinear absorption and photoluminescence.} 
CsPbBr$_3$ perovskite exhibits a very strong photoluminescence (PL) via efficient nonlinear absorption where the corresponding three-photon and two-photon absorption coefficients were measured to be \red{$\approx$}~0.14 $\pm$ 0.03~cm$^3$/GW$^2$ at 1200~nm~\cite{clark2016polarization} and \red{$\approx$}~4.57~cm/GW at 800~nm~\cite{liu2018robust}, respectively. These values are \red{up to} three times larger than those known for conventional semiconductors with the similar band gaps such as GaP and CdS~\cite{nathan1985review}. It should be noticed that 3PA-active materials are not necessarily luminescent especially at room temperature; for example, the 3PA-induced PL in CdS is sufficiently bright only at cryogenic temperatures. Moreover, PL from single crystal CsPbBr$_3$ can be linearly polarized~\cite{clark2016polarization}. Strong third-order nonlinearity was also observed in MAPbI$_3$ thin films, where nonlinear refractive index is \red{$\approx$}3.74 $\times$10$^{−15}$ m$^2$/W at 1064~nm~\cite{zhang2016nonlinear}, being three orders of magnitude larger than that of silicon. Nonlinear processes in halide perovskites were shown to be suitable for laser  mode-locking~\cite{walters2015two, zhang2016nonlinear}, and for efficient optical pumping of nanolasers~\cite{liu2018robust}.

\begin{figure}[h!]
 \centering
\includegraphics[width=0.99\linewidth]{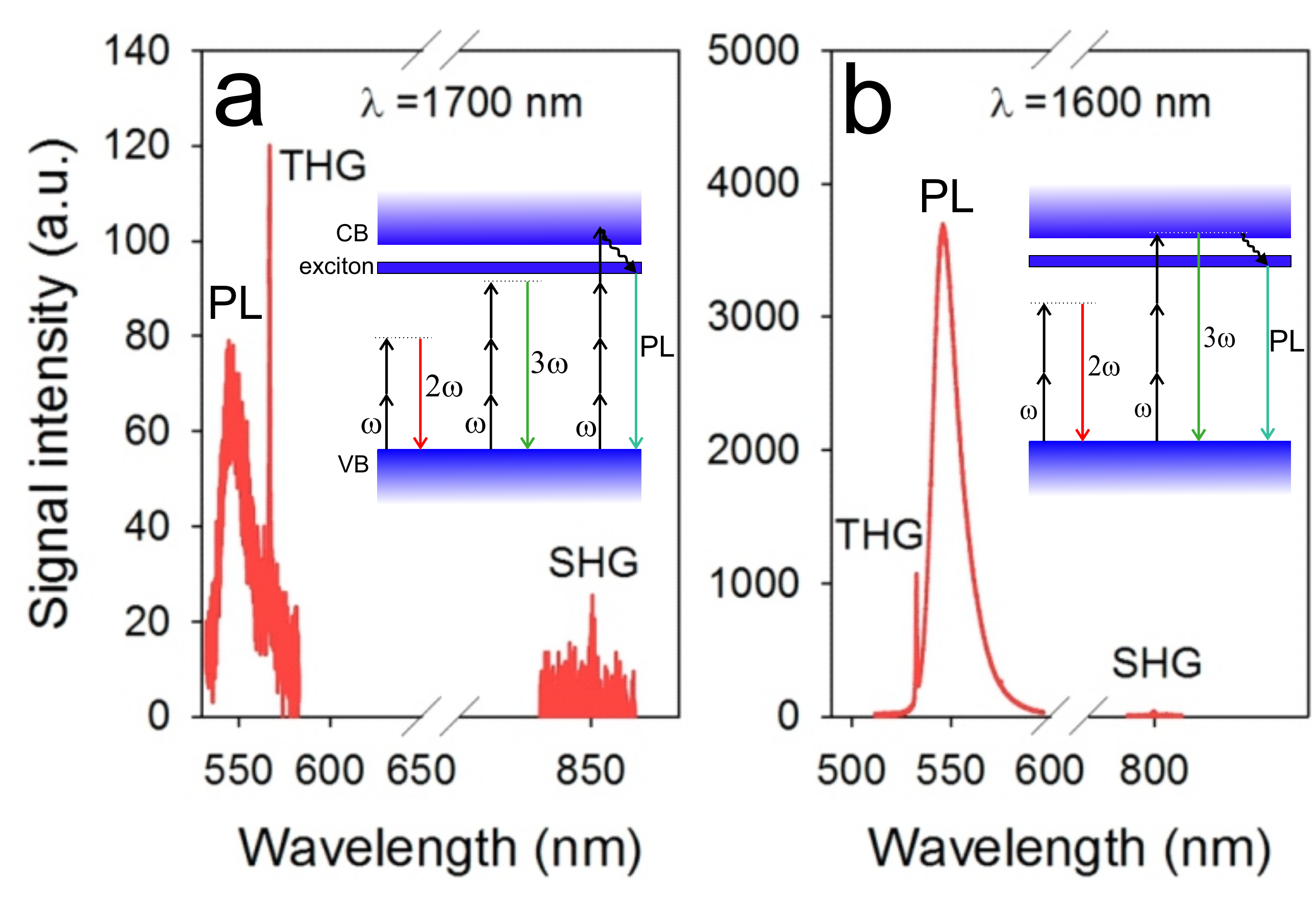}
\caption{\textbf{Harmonics generation from halide perovskites.} Spectra of photoluminescence (PL), second-harmonic (SHG) and third-harmonic generation (THG) from CsPbBr$_3$ at different excitation wavelengths (of a femtosecond laser): (a) 1700~nm and (b) 1600~nm. \red{Insets schematically show level diagrams and processes of SHG, THG, and PL at corresponding pulsed laser optical excitation. Adapted with permission from~\cite{clark2016polarization}. Copyright 2018 by the American Physical Society.}}
\label{harmonics}
\end{figure}

\textbf{Generation of optical harmonics.}
Despite the fact that perovskites are known as nonlinear optical materials for the second-harmonics generation (SHG) (e.g., for BaTiO$_3$), the most popular halide perovskites (CsPbX$_3$ or MAPbX$_3$) are less effective for SHG, as shown in Fig.~\ref{harmonics}. The reason is a centrosymmetric crystalline structure of these perovskites (e.g. cubic), and even nonsymmetrical organic molecula (e.g. MA=CH$_3$NH$_3$) does not encrease much second-order nonlinearity~\cite{sharada2016ch, yang2018lasing}. In this regard, one has to change the composition to make halide perovskites noncentrosymmetric. Giant SHG has been observed from the halide perovskites where led was replced by Ge (AGeX$_3$), exhibiting  a large second-order nonlinear response ($\chi^{(2)}$(CsGeI$_3$)=125.3 pm/V, The value $\chi^{(2)}$(MAGeI$_3$)=161.0~pm/V) for MAGeI$_3$~\cite{stoumpos2015hybrid} is comparable with conventional strongly nonlinear materials such as GaAs with $\chi^{(2)}$(GaAs)=750 pm/V. However, creation of lead-based halide perovskites with huge values of $\chi^{(2)}$ is still challenging. \red{For example, Fig.~\ref{harmonics} shows that the one of the most widespread compounds CsPbBr$_3$
is essentially centrosymmetric as clearly evidenced by much
stronger THG over SHG, even when THG is being significantly reabsorbed by the perovskite. Very strong THG was observed from Ruddlesden-Popper layered perovskites as well, where a maximum effective third-order susceptibility $\chi^{(3)}$ was around 1.12$\times$10$^{-17}$~m$^2$V$^{-2}$~\cite{abdelwahab2017highly}.}

\textbf{Terahertz emission.} 
Strong THz emission from thin films of MAPbI$_3$  has been observed in Ref.~\cite{guzelturk2018terahertz} with the amplitudes comparable to those of single-crystal semiconductors, and high optical-to-THz conversion efficiencies was calculated  and observed ($\approx$10$^{-3}$). It was proven that the emitted THz fields are generated by an ultrafast transient photocurrent normal to the film surface, arising from a difference in respective diffusivities of electrons and holes, this phenomenon is known as the photo-Dember effect~\cite{gu2002study}. The authors used polycrystalline MAPbI$_3$ films on glass substrates prepared by both spin coating and thermal evaporation techniques, which resulted in similar THz emission properties. It worth noting that adding bromine to the perovskite composition leads to a blue shift of the THz emission. 

\subsection{Dynamic tunability}

\textbf{Photo-generated carriers.}
Because polycrystalline perovskites have sub-nanosecond photocarrier recombination lifetimes, switching between resonances can occur on ultrafast timescales. The photocarrier recombination kinetics have been shown to be modeled by the rate equation given by
\begin{equation}
\frac{dN}{dt}= -k_3N^3-k_2N^2-k_1N
\end{equation}
where $k_1$, $k_2$, and $k_3$ are the recombination rate constants and $N$ is the photocarrier density~\cite{noh2013chemical}. For low optical
fluences, the photocarrier dynamics is dominated by a monomolecular decay (−$k_1N$), whereas at high pump fluences, the recombination dynamics is dominated by electron-hole bimolecular recombination (−$k_2N^2$) and Auger recombination (−$k_3N^3$), respectively~\cite{noh2013chemical}. In Ref.~\cite{chanana2018ultrafast}, the monomolecular recombination lifetime ($\tau_{rec}$=k$_1^{-1}$) for MAPbI$_3$ $\approx$~7.8~ns and for MAPbBr$_3$ $\approx$~0.5~ns  at low optical fluence have been measured. At higher fluences, the other contributions are estimated to be: $k_3N^2\approx$~2.5$\times$10$^{10}$s$^{-1}$ and $k_2N\approx$~3.1$\times$10$^9$s$^{-1}$ for MAPbI$_3$ (fs-laser 800-nm pump wavelength with fluence 310~\red{$\mu$J/cm$^2$}); $k_3N^2\approx$~5.5$\times$10$^{10}$s$^{-1}$ and $k_2N\approx$~8.3$\times$10$^9$s$^{-1}$ for MAPbBr$_3$ (fs-laser 400-nm pump wavelength with fluence 105~\red{$\mu$J/cm$^2$}). Close photo-generated dynamics was observed as well~\cite{manjappa2017hybrid}.
The density of photoexcited carriers is usually observed up to 10$^{20}$cm$^{-1}$ that is enough to modulate the tramsmittance of perovskite films up to 10--30~\% level for a signal beam at THz frequencies~\cite{chanana2018ultrafast}, and 0.1~\% level for the visible range~\cite{sheng2015exciton}.

\textbf{Light-induced migration of ions.}
As mentioned above, substantial tuning of optical properties of lead halide perovskites  can be achieved by the synthesis of heterohalogenide APbCl$_{3-x}$Br$_x$ or APbBr$_{3-x}$I$_x$ ($0<x<3$) structures. However, thin films of MAPbBr$_{3-x}$I$_x$ demonstrate reversible (recovered in darkness) light-induced ions migration and the formation of bromine-rich and iodine-rich domains at extremely low intensities $<$100~mW/cm$^2$~\cite{hoke2015reversible}. The latter behave as trap centers for electron-hole recombination. For that reason, the attenuation of emission at 1.85~eV and the development of a new red-shifted PL signal at 1.68~eV growing in intensity under constant illumination did occur in one minute. \red{Also, the segregation effect caused reversible variation by more than three orders of magnitude of absorption coefficient near the bandgap edge}~\cite{hoke2015reversible}. Draguta {\it et al.} demonstrated the influence of pump power on the rate of phase separation in MAPb(Br$_{0.5}$I$_{0.5}$)$_3$~\cite{draguta2017rationalizing}. The segregation effect is prospective for intensity dependent dynamic spectral tuning of perovskite \red{nanostructures}, which is unusual mechanism for conventional materials of nanophotonics. However, despite giant optical changes at low intensities, a typical timescale of this mechanism is around 0.1--10~s, being many orders of magnitude lower compared with commonly employed Kerr-like nonlinearities~\cite{makarov2017light}.

\textbf{Anion exchange methods.}
A good alternative to fabrication of mixed-halide lead perovskites is modification of the monohalide structures by anion-exchange method. The substitution of halogen atoms in the solid crystal lattice can be conducted in three different ways including reaction of prepared APbX$_3$ nanocrystals with organic salts (MA-X, ODA-X, OLAM-X, TBA-X) in solution~\cite{akkerman2015tuning}, material rehalogenation in the presence of X$_2$ or HX (X = Cl, Br, I) gas~\cite{solis2015post, chen2016pseudomorphic, he2017multi, he2017patterning} and solid state anion interdiffusion detected in osculating nano- and microobjects~\cite{pan2018visualization}.
The first method allows for formation of the heterohalide perovskites with optical characteristics determined by concentration of the organic precursors taken for the reaction. On the contrary, the gas assisted method allows for slow and precise changing of photophysical properties by changing treatment time from a few minutes to tens of hours at room temperature can be tracked with the help of a fluorescent microscope.   

The method of gas assisted anion-exchange is extremely prospective for \textit{in-situ} modification of halide perovskite nanostructures integrated to various photonic designs. Indeed, reversible tunability of the nanostructures optical properties over entire visible range might revolutionize the field of reconfigurable nanophotonics~\cite{makarov2017light}. 
    
\textbf{Temperature effects.}
Optical properties of halide perovskites depend strongly on temperature. The main reason of temperature-driven changes of the PL spectra in organo-lead halide perovskites originates from the transitions between their structural phases possessing different optical features. 

MAPbI$_3$ has an orthorhombic (Orth) phase below 162~K, a tetragonal (Tet) phase in the 162--327~K range, and a cubic one above 327~K~\cite{poglitsch1987dynamic, stoumpos131}. The low-temperature Orth phase measured at 15~K was shown to exhibit two emission peaks located at 1.574 and 1.649~eV that experienced a blue-shift with temperatures rise~\cite{dar2016origin, chen2017origin}. The high-energy emission peak vanishing at 120~K was assigned to Orth domains with molecularly disordered methylammonium cations while the unusual blue-shift contradicting with the Varshni behavior of semiconductors~\cite{varshni1967temperature} was explained by the stabilization of the valence band maximum~\cite{dar2016origin}. The single emission band undergoing a similar systematic blue-shift from 1.569 up to 1.601~eV in the 150--300~K range was attributed to the Tet phase. This signal had gradually been transformed to luminescence of the cubic phase featured with slightly lower energy (1.62 eV) at 400 K~\cite{chen2017origin}. 

Similar temperature dependence is revealed for green-light emission of MAPbBr$_3$ perovskites, however, the replacement of MA$^+$ by FA$^+$ in the crystalline lattice lead to the disappearance of the low-temperature high-energy PL peak, which confirm that the dual emission is related to the nature of the organic cation~\cite{dar2016origin}. Moreover, such a replacement results in temperature-history dependent PL for FAPbI$_3$ as compared with that of MAPbI$_3$~\cite{chen2017origin}. Considerably different luminescent behaviors for FAPbI$_3$ thin films having initially hexagonal (Hex) and cubic phases were demonstrated while the samples were cooling down. The Hex perovskite film demonstrate broad and high-energy irradiation spectra at $\sim$1.85~eV corresponding to intermediate- and low-temperature hexagonal phases, whereas the cubic perovskite films show narrow-band emission at $\sim$1.5~eV assigned \red{to both intermediate- and low-temperature Tet phases}.

Among the class of lead halide perovskites, the most temperature-sensitive optical characteristics are observed for CsPbI$_3$. Its effectively luminescent and highly absorbing (1.73~eV bandgap) cubic phase exists at temperatures higher than 583~K and transforms to nonemissive and semitransparent orthorhombic phase (2.82~eV bandgap) below 583 K that hinders this material utilization for light-convertion and light-emission applications without careful processing control and development of a low-temperature phase transition~\cite{eperon2015inorganic}.

\section{Perovskite nanostructures}

In this section, we discuss nanostructures made of halide perovskites with typical dimensions much larger than the Bohr radius of exciton (e.g., $r_B \approx$2--6~nm for CsPb(Cl,Br,I)$_3$ perovskites~\cite{protesescu2015}), but less or comparable with the wavelength of incident light. In this case, the confinement of excitons is negligible, whereas light localization can be made strong due to the \red{excitation} of optical modes of different origin and for different geometries. 

\subsection{Nanoparticles}

\textbf{Fabrication methods.} Resonant perovskite nanoparticles usually acquire spherical or cubic shapes with a typical diameter in the range 200-1000~nm. This brings novel physics of meta-optics associated with the low-order resonant geometric Mie-type modes.

There are knwon several ways to fabricate perovskite nanoparticles with submicron sizes, such as laser ablation~\cite{tiguntseva2018light}, CVD~\cite{tang2017single}, or chemical synthesis~\cite{huang2016colloidal}. Laser ablation represents the fast physical method to obtain single perovskite nanoparticles from perovskite films. Here nanoparticles takes a quasi-spherical shape with a broad size distribution. To achieve better spherical forms of CsPbX$_3$ nanoparticles, Tang~\cite{tang2017single} used the CVD technique resulting in broad size distribution approximately 0.5 to 10~$\mu$m. To prepare perovskite NPs antisolvent precipitation~\cite{jia2017facile} or low-temperature growth~\cite{liu2018robust} syntheses are used. Chemical methods allow to variate shape and size and present usually a one step synthesis procedure, where size of obtained nanoparticles depends on the perovskite solution concentration, antisolvent \red{choice}~\cite{paek2017nano}, additional surfactants~\cite{huang2015emulsion, gonzalez2016luminescence}, and solution concentration. Manipulating these parameters, it is possible to obtain perovskite nanoparticles with the sizes ranging from a few \red{nanometers}~\cite{schmidt2014nontemplate} to a micrometer~\cite{jia2017facile} or submicron size with a cubic shape~\cite{liu2018robust}.

\begin{figure*}
 \centering
\includegraphics[width=0.85\linewidth]{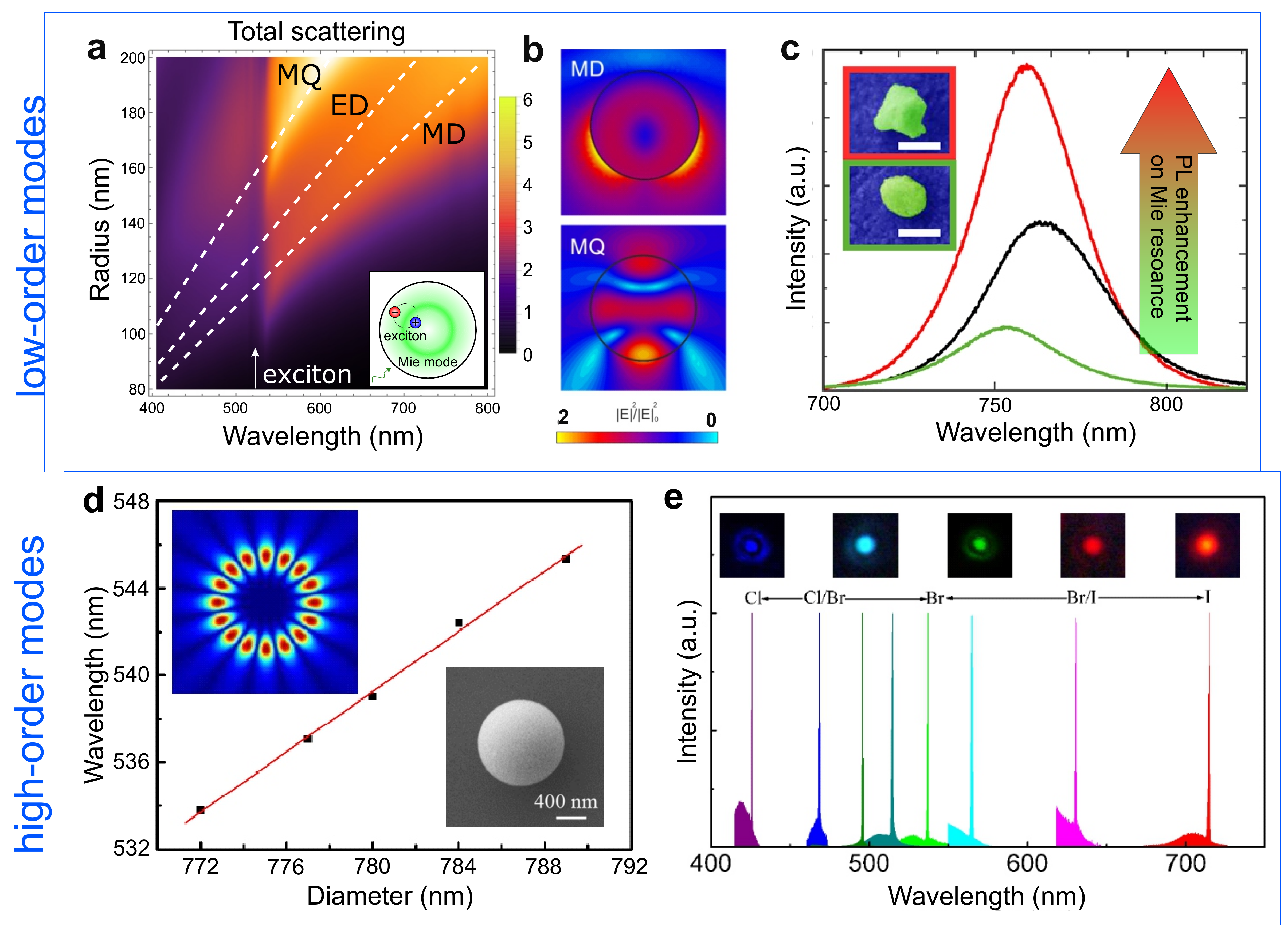}
\caption{\textbf{Resonant halide-perovskite nanoparticles.} (a) \red{Analytically calculated total scattering efficiency $Q_{scat}$ of a spherical MAPbBr$_3$ nanoparticle in vacuum. Overlaid lines depict approximate dispersion of the electric dipole, magnetic dipole, and magnetic quadrupole resonances of the particle. The arrow indicates position of MAPbBr$_3$ exciton.} (b)  Calculated near-field distributions for \red{single MAPbI$_3$ nanoparticles supporting the magnetic dipolar and quadrupolar resonances at emission wavelength}.\red{Adapted with permission from~\cite{tiguntseva2018light}. Copyright 2018 American Chemical Society.} (c) PL enhancement for a nanoparticle at magnetic quadrupole resonance \red{(red curve)}, as compared with thin a film  \red{(black curve)} and non-resonant nanoparticle  \red{(green curve)}. \red{Adapted with permission from~\cite{tiguntseva2018light}. Copyright 2018 American Chemical Society.} (d) Lasing modes in CVD-grown CsPbBr$_3$ submicron particles of different sizes.\red{Reprinted with permission from~\cite{tang2017single}. Copyright 2018 American Chemical Society.} (e) Lasing from perovskite submicron \red{CsPbX$_3$ spheres with different halogens X: Cl,Br, I and mixed ones}. \red{Reprinted with permission from~\cite{tang2017single}. Copyright 2018 American Chemical Society.}}
\label{NPs}
\end{figure*}

\textbf{Resonant properties.} Recently, resonant dielectric nanoparticles with high refractive index have attracted a lot of attention in photonics as an alternative approach to achieve a strong resonant response and subwavelength light localization~\cite{kuznetsov2016optically}. Optical resonances in a spherical dielectric particle can be described analytically by the Mie theory~\cite{mie1908beitrage}, while numerical methods are required to solve the Maxwell equations for cylindrical, cubical, conical, and other shapes of nanoparticles~\cite{evlyukhin2011multipole}.   
The \textit{first resonance} of a spherical dielectric nanoparticle is the magnetic dipole resonance, and it occurs when the wavelength of light ($\lambda$) inside a spherical particle with the refractive index (\textit{n}) becomes comparable to particle's diameters, (\textit{D})~\cite{Bohren1998small}: 
\begin{equation}
D_{\rm res}^{(1)} \approx \lambda / n  .
\label{MD}
\end{equation}
Under this condition, the polarization of the electric field is antiparallel at the opposite boundaries of the nanoparticle, which results in coupling to the circulating displacement currents with the magnetic field oscillating in the center~\cite{kuznetsov2012magnetic}. Next low-order multipoles (quadrupole, octupole, etc.) are useful for enhancing light-matter interaction and lead to large near-field densities inside the dielectric particle~\cite{kapitanova2017giant}, boosting nonlinear optical response~\cite{shcherbakov2014enhanced, makarov2017efficient}, Raman signal~\cite{RamanNanoscale}, as well as photoluminescence~\cite{rutckaia2017quantum, zalogina2018purcell} in silicon-based nanostructures. Also, interference between electric and magnetic modes results in highly reconfigurable scattering power patterns~\cite{kerker1983electromagnetic, fu2013directional, makarov2015tuning}. 

According to the data from Fig.~\ref{Optics}(b,c) and Eq.~\ref{MD}, in a spherical halide perovskite (e.g. MAPbBr$_3$) nanoparticle the first Mie resonance (magnetic dipole) can occur near its emission line at the diameters D$_{res}\approx$200--250~nm, whereas larger sizes \red{allow} for excitation of higher-order modes [see Fig.\ref{NPs}(a)]~\cite{tiguntseva2018light}. Similar behavior is observed experimentally for MAPbI$_3$ nanoparticles~\cite{tiguntseva2018light}, where the near-field structures at the magnetic dipole and quadrupole resonances are shown in Fig.~\ref{NPs}b. 

Furthermore, since they possess pronounced exciton state with binding energy much more than 25~meV~\cite{tanaka2003comparative} (\red{Sec.~2.2}), it is expected that coherent coupling of Mie modes of the particle to the excitons of perovskite~\cite{platts2009whispering} will result in a pronounced Fano resonance at room temperature. Calculated scattering cross section spectra of spherical nanoparticles versus particle radius presented in \red{Fig.\ref{NPs}}a reveals an asymmetric behavior inherent for Fano-like dip close to the exciton resonance of MAPbBr$_3$ perovskite at 539~nm. This knowledge is crucial for the analysis of transmittance spectra for the films made of perovskite nanoparticles, where the dip around exciton spectral line has to be attributed to the Fano resonance rather than reduced absorption in the material.

Photoluminescence at the Mie resonances is enhanced because the spontaneous emission is accelerated due to the Purcell effect~\cite{zambrana2015purcell}. Namely,  according to the Fermi`s Golden rule the spontaneous emission rate $\Gamma$ of a point dipole emitter in a cavity relative to the emission rate $\Gamma_0$ in free space depends on the interaction strength of the dipole emitter with the cavity mode field~\cite{englund2005controlling},
\begin{equation}
F =\frac{P_t}{P_0}= \frac{\Gamma}{\Gamma_0}= F_p \frac{|\textbf{u}(\textbf{r}_{em})\cdot\textbf{d}|^2}{|\textbf{d}|^2}\frac{1}{1+(\frac{2Q}{\hbar\omega_c})^2(\hbar\omega_0-\hbar\omega_c)^2},
\label{Purcell}
\end{equation}
where $P_t$ is the total power dissipated by the dipole, $1+(2Q/\hbar\omega_c)^2(\hbar\omega_0-\hbar\omega_c)^2$ describes the spectral mismatch between the
emitter oscillation frequency $\omega_0$ and mode frequency $\omega_c$, 
$Q=\omega_c/\gamma$ is the quality factor of the
mode, $\textbf{u}(\textbf{r}_{em})$ is the mode amplitude at the position of the dipole \textbf{\textit{r}}$_{em}$ and \textbf{\textit{d}} is the dipole moment, F$_p$ is the maximal value of Purcell factor.  The Purcell factor reaches then its maximal value $F = F_p$ which in case of closed resonators has the simple form, 
\begin{equation}
F_p = (6\pi c^3/n^3\omega_c) (Q/V_{mode}),
\end{equation}
where \textit{n} is the refractive index and $V_{mode}$ is the mode volume. Here, it becomes clear that with increase of the $Q$-factor, one can expect both acceleration of spontaneous emission rate $\Gamma$, as well as the emitted power.  However, optical losses in perovskites at the exciton peak [see Fig.~\ref{Optics}(c)] and increasing mode volume with a growth of particle's size reduces an averaged emitted power, being maximum for the magnetic quadrupole \red{resonance}~\cite{tiguntseva2018light}.

For larger spherical particles, higher Mie-type modes can support amplified spontaneous emission and even lasing, provided the $Q$-factor and gain are high enough to overcome losses. In general, the threshold value of the $Q$ factor is inversely proportional to the optical gain $g$, which is higher than 10$^2$cm$^{-1}$ for various halide perovskites (see Table II).
The single-mode laser with a very narrow linewidth ($\approx$0.09~nm) is realized in a perovskite submicron spherical cavity at low threshold (0.42~$\mu$J/cm$^2$) with high cavity $Q$ factor of lasing line ($\sim$6100)~\cite{tang2017single}. Additionally, combining \red{composition} modulation and size control, high quality single-mode lasing can be extended to the whole visible spectra range, as shown in Figs.~\ref{NPs}(d,\red{e}).

Cuboid CsPbBr$_3$ nanoparticles smaller than 500~nm also demonstrated lasing behavior~\cite{liu2018robust}. Single-mode lasing was achieved with the thresholds of 40.2~$\mu$J/cm$^2$ and 374~$\mu$J/cm$^2$ for single-photon and two-photon fs-laser pump, respectively, with determined gain around 500~cm$^{-1}$. The $Q$-factor of the lasing was estimated to be around $Q_{las}$~$\approx$~2$\times$10$^3$ with the line width of 0.26~nm at the wavelength of 540~nm. Remarkably, the fast exciton relaxation in the lasing regime results in 22-ps duration of the emitted pulses from such nanolasers.

Thus, we notice that the resonant halide-perovskite nanoparticles are rapidly becoming versatile building blocks for various nanophotonic applications, and they demonstrate strong coherent and incoherent light emission at the wavelengths covering the entire visible spectrum.

\subsection{Nanowires and nanoplates}

\begin{figure}[hb!]
\includegraphics[width=0.99\linewidth]{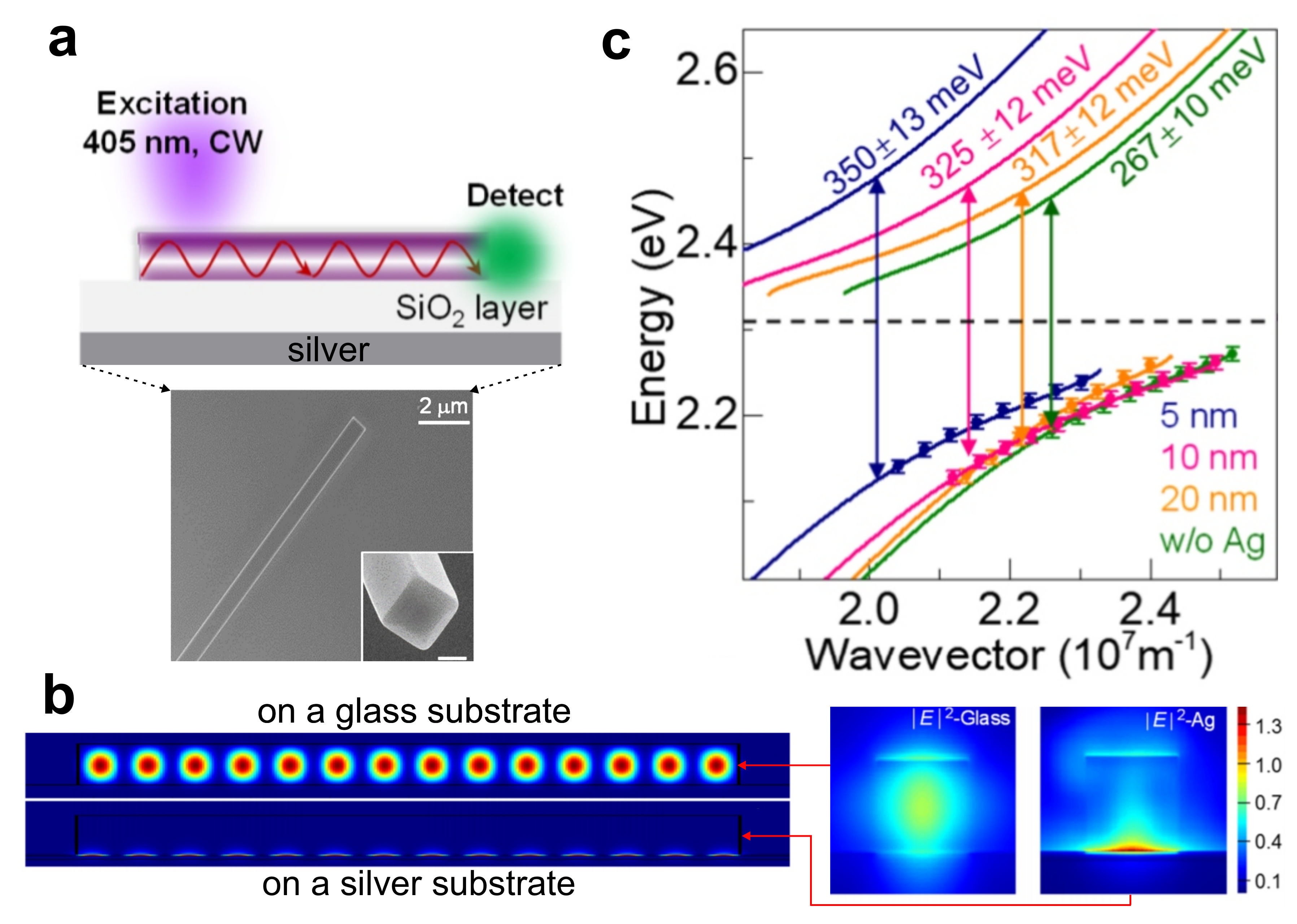}
\caption{\textbf{Strong coupling of excitons to Fabry-Perot modes in perovskite nanowires.} (a) Schematic of a perovskite nanowire photoexcitation and detection. Inset: electron image of the MAPbBr$_3$ nanowire. (b) Calculated electric near-field distribution in the nanowire placed on glass and silver substrates. (c) Dispersion curves for exciton-polaritons in the nanowire on a silver substrate with different SiO$_2$ spacers (5, 10, 20~nm) and without silver. Solid lines represent the fittings by formulas similar to Eq. (10). \red{Adapted with permission from~\cite{shang2018surface}. Copyright 2018 American Chemical Society.}}
\label{NW}
\end{figure}

\textbf{Fabrication methods.} Nanowires and nanorods are structures with at least one dimension going to the nanoscale ($<$500~nm), while other two can range from a few microns up to several hundred micrometers \red{[see an example in Fig.~\ref{NW}(a)]}. Nanowires fabricated from halide perovskite are excellent objects for wavequided photoluminescence and even lasing, as shown schematically in Fig.~\ref{NW}(a). There are two approaches to synthesize perovskite nanowires including wet-chemistry~\cite{zhang2018strong, evans2018continuous} and CVD~\cite{xing2015vapor, du2017strong}. Wet chemistry can be realised by exploiting methods such as nanowire precipitation by slow diffusion of an antisolvent into a perovskite solution, a growth from solution regions restrained by a patterned \red{polymer (e.g. PDMS)} template or growth of nanolasers as a dense forest on a surfactant-covered (e.g. PEDOT:PSS) substrate with a thin layer of PbI$_2$. 
\red{As was demonstrated by Zhu {\it et al.}~\cite{zhu2015lead}, in the two-step reaction lead acetate is first turned into PbI$_4$ ions, which subsequently react with the MAX solution to form single-crystal nanowires on a substrate with flat-end facets and lengths up to 20~$\mu$m.}

\red{In the CVD process, perovskite is heated above its sublimation point in an inert atmosphere. The CVD approach can be realised by means of evaporation of PbX$_2$ and AX salts in the stream of pure Ar or N$_2$ gases. The vapor resublimates onto a substrate, which is kept at a lower temperature than the perovskite source. The improved crystal quality and stability was achieved in the CVD-grown nanorods~\cite{zhou2016vapor}.} Since the fabrication occurs in an inert gas atmosphere and the individual rods are separated by alumina walls, water diffusion between the rods is suppressed and crystals can be formed without air or water inclusions~\cite{waleed171}.

\textbf{Resonant properties.}
The nanowire interfaces are flat crystal grain boundaries, and they can act as Fabry-Perot optical cavities with two end mirrors, while the light is guided along the axial waveguide formed by the other surfaces~\cite{zhu2015lead,yan09}. In general, an infinite dielectric cylindrical nanowire operates as an optical waveguide and supports the transverse electric (TE$_{0m}$) and transverse magnetic (TM$_{0m}$) modes, which have only three field components, and also hybrid modes (HE$_{nm}$ and EH$_{nm}$) which are described by six field components. In free-standing nanowire, all modes, except HE$_{11}$, are characterized by low-frequency cutoffs~\cite{collin1960field}. A typical \red{mode structure} in MAPbBr$_3$ nanowire placed on a silica substrate is shown in Fig.~\ref{NW}(b).

Reflections between the end surfaces of a nanowire create Fabry-Perot resonances, while the light is guided along the axial waveguide formed by the other surfaces. Tuning the length, \red{nanowire} medium, and surrounding medium influence the \red{resonant properties}. For a Fabry-Perot cavity of length $L$, the mode spacing ($\Delta\lambda$ ) at $\lambda$ is given by~\cite{o2007microcavity}
\begin{equation}
\Delta\lambda = (\lambda^2/2L)[n - \lambda(dn/d\lambda)]^{-1}
\end{equation}
where \textit{n} is the refractive index and \red{$dn/d\lambda$} is the dispersion relation. Quality factor $Q$ of the nanowire is another important parameter being expressed as~\cite{maslov2003reflection}
\begin{equation}
Q = -Lk_{||}/ln|r_1r_2|
\end{equation}
where $r_1$ and $r_2$ the reflection coefficients at each facet, $L$ is the nanowire length, and $k_{||}$ is wavenumber along the nanowire axis. Typical values of the $Q$-factor for halide perovskite nanowires (with the length ranging from a few microns up to tens of microns) are in the range of \textit{Q}$\sim$50--1000.

Remarkably, nanowires allow for confinement of light into the effective mode volume \red{(considering that excitons are distributed throughout the entire crystal, not just at the position of maximum field)}, which can be calculated numerically as:
\begin{equation}
V_{eff}=A\cdot L\frac{\int_{A_s}{\epsilon(x,y)|E(x,y)|^2dxdy}}{\int_{A}{\epsilon(x,y)|E(x,y)|^2dxdy}}
\end{equation}
where $A_s$ is the simulation area, \textit{A} is the geometric cross-section area of a nanowire, and \textit{L} is the length of a nanowire. Typical values of the effective mode volumes for high-index dielectric nanowires placed on low-index dielectric substrates are diffraction limited, whereas it can be significantly reduced by placing the nanowire on metal supporting surface plasmon modes~\cite{landreman2016fabry}. 

\begin{figure*}
\includegraphics[width=0.85\linewidth]{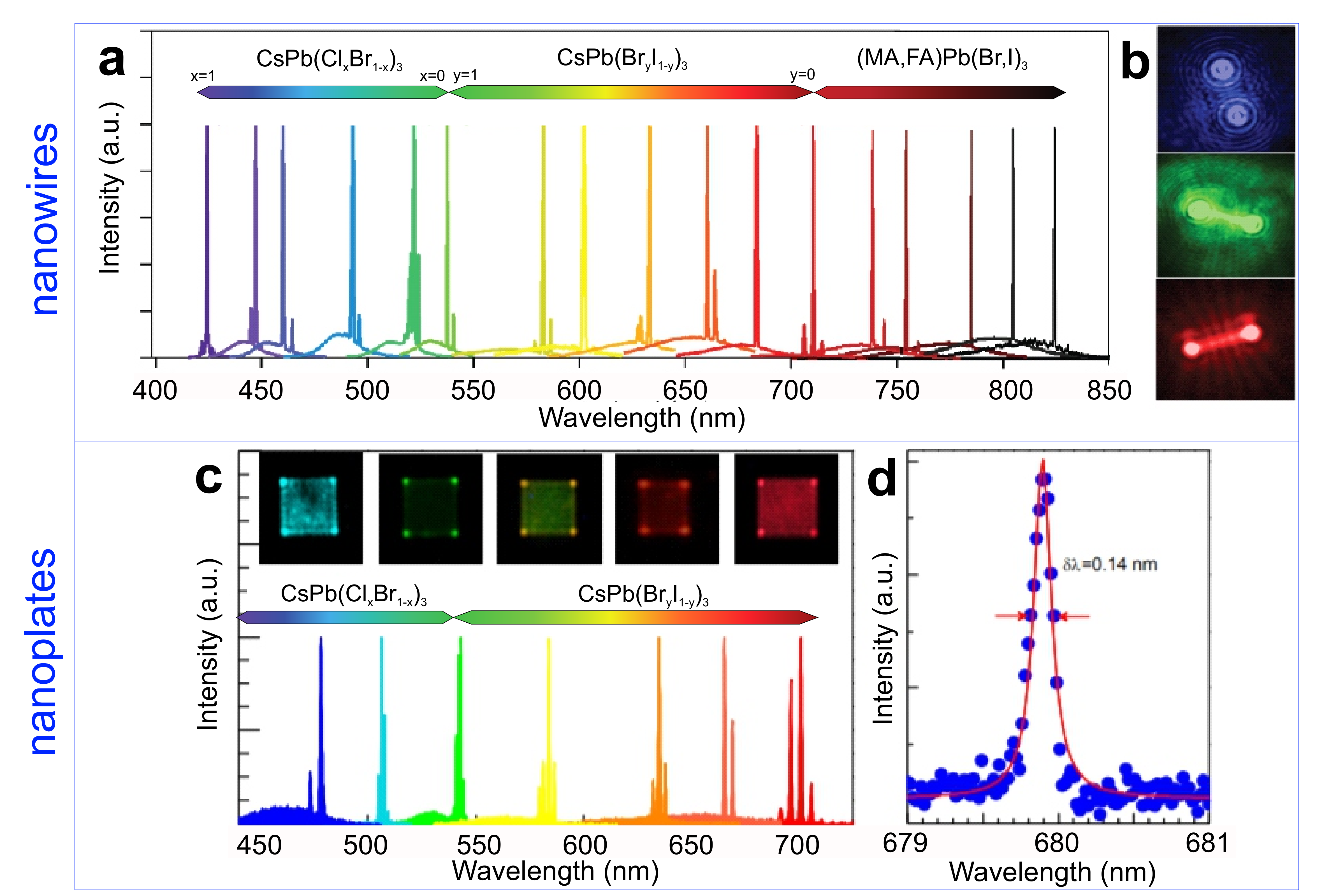}
\caption{\footnotesize{\textbf{Perovskite nanowires and nanoplates.}} (a) Lasing spectra from perovskite nanowires of various composition and length in range 5--20$\mu$m. (b) Optical images in the lasing mode operation. Adapted from Refs.~\cite{zhu2015lead,fu2016nanowire}. (c) Lasing spectra from perovskite nanoplates of various composition. Inset: optical images in the lasing mode operation. (d) Typical lasing spectrum of a nanoplate corresponding to Q$_{las}\approx$4850. Adapted from Ref.~\cite{zhang2016high}.}
\label{NWlas}
\end{figure*}

Coupling between the exciton resonance and light is enhanced in a nanowire due to the reduced mode volume of photons and the cavity-enhanced oscillator strength described by the relation
\begin{equation}
\Omega \sim \sqrt[]{f/V_{eff}},
\end{equation}
where $\Omega$ is the vacuum Rabi splitting, \textit{f} is the oscillator strength, and V$_{eff}$ is the effective mode volume. Strong coupling prevails when the vacuum Rabi frequency ($\Omega$/$\hbar$) becomes larger than the exciton dephasing rate ($\Gamma$), $\Omega$ $>$ $\hbar\Gamma$. In other words, the photon lifetime as determined by cavity $Q$-factor should be long enough to allow for the Rabi oscillation to occur. In this regard, perovskite nanowires with high oscillator strength ($f \sim 0.1-0.2$~exciton$^{-1}$~\cite{du2017strong}) supporting high-$Q$ Fabry-Perot modes should demonstrate strong coupling of excitons with the cavity. The strong coupling regime results in splitting of the dispersion for optical modes in a nanowire around the exciton line to two branches, the upper (+) and lower(-), for the polariton dispersion:
\begin{equation}
E_{\pm}=\frac{1}{2}(E_{ex}+E_{ph})\pm\frac{1}{2}\left((E_{ex}-E_{ph})^2+\Omega^2\right)^{1/2}
\end{equation}
where 
\[E_{ph}=c\left(\hbar^2k_{||}^2+m_0^2c^2\right)^{1/2}, \;\; m_0=(\pi\hbar/c)\left(a^{-2}+b^{-2}\right)^{1/2}, 
\]
and \textit{a} and \textit{b} are the cross-sectional widths of the nanowire. The wavevector $k_{||}$ is \red{determined as $k_{||}=k_0 + j\pi/L$, where $j$ is integer number and $k_0$ is the wavenumber of the zeroth mode}. Importantly, halide perovskites are always high-loss materials for the photon energies higher than the exciton state, owing to a strong interband absorption, remaining lower polariton branch only. The example of such a Rabi splitting is presented in Fig.~\ref{NW}(c), where experimentally observed Fabry-Perot modes in MAPbBr$_3$ perovskite nanowires are lying on the polaritonic dispersion curve.

\begin{table}[]
\centering
\caption{Rabi splitting in nanowires}
\label{Rabi}
\begin{tabular}{|l|l|l|l|}
\hline
Material & Structure     & Rabi splitting & Refs.               \\ \hline
ZnO      & Nanowire on sapphire  & 160 meV        & \cite{van2006exciton}      \\ \hline
GaN      & Nanowire in Bragg-mirror  & 48 meV         & \cite{das2011room}         \\ \hline
CdS      & Nanowire on SiO$_2$     & 200 meV        & \cite{van2011one}          \\ \hline
CsPbBr$_3$  & Nanowire on sapphire  & 200 meV        & \cite{evans2018continuous} \\
         & Nanowire on sapphire     & 146 meV        & \cite{wang2018high}\\
         & Nanowire on Si/SiO$_2$      & 650 meV        & \cite{du2017strong}        \\ \hline
MAPbBr$_3$  & Nanowire on indium tin oxide & 390 meV        & \cite{zhang2018strong}     \\ 
  & Nanowire on metal     & 560 meV        & \cite{shang2018surface}     \\ \hline
  CsPbI$_3$  & Nanowire on sapphire & 103 meV        & \cite{wang2018high}      \\ \hline
  CsPbCl$_3$  & Nanowire on sapphire & 210 meV        & \cite{wang2018high}      \\ \hline
\end{tabular}
\end{table}

Table~\ref{Rabi} compares the Rabi splitting energies for nanowires made of different halide perovskites with those for conventional semiconductors (ZnO, GaN, CdS). A large Rabi splitting indicates a large oscillator strength, which is related to the exciton binding energy. In perovskites, while an above gap excitation creates predominantly charge carriers at room temperatures, there is a strong excitonic resonance in absorption and fluorescence emission, with an exciton binding energies exceeding the limit 25~meV. As a result, very high splitting values ($\Omega$$\approx$200--650~meV) have been reported recently~\cite{du2017strong,shang2018surface,zhang2018strong,evans2018continuous}. \red{Chemical} compositions dependence of exciton-photon coupling systematically was studied in crystalline CsPbX3 perovskite nanowires by Wang et al.~\cite{wang2018high}, revealing two-fold decrease of Rabi splitting energy from CsPbCl$_3$ to CsPbI$_3$ caused by reducing of exciton binding energy \red{(see Sec.~2.2)}. Remarkably, additional coupling of nanowire's modes with surface plasmons can be employed for the effective mode volume V$_{eff}$ compressing [see Fig.~\ref{NW}(b)], resulting in a significant increase of the Rabi splitting energy from 270~meV to 350~meV~\cite{shang2018surface}.

\begin{figure*}
\noindent \includegraphics[width=0.9\linewidth]{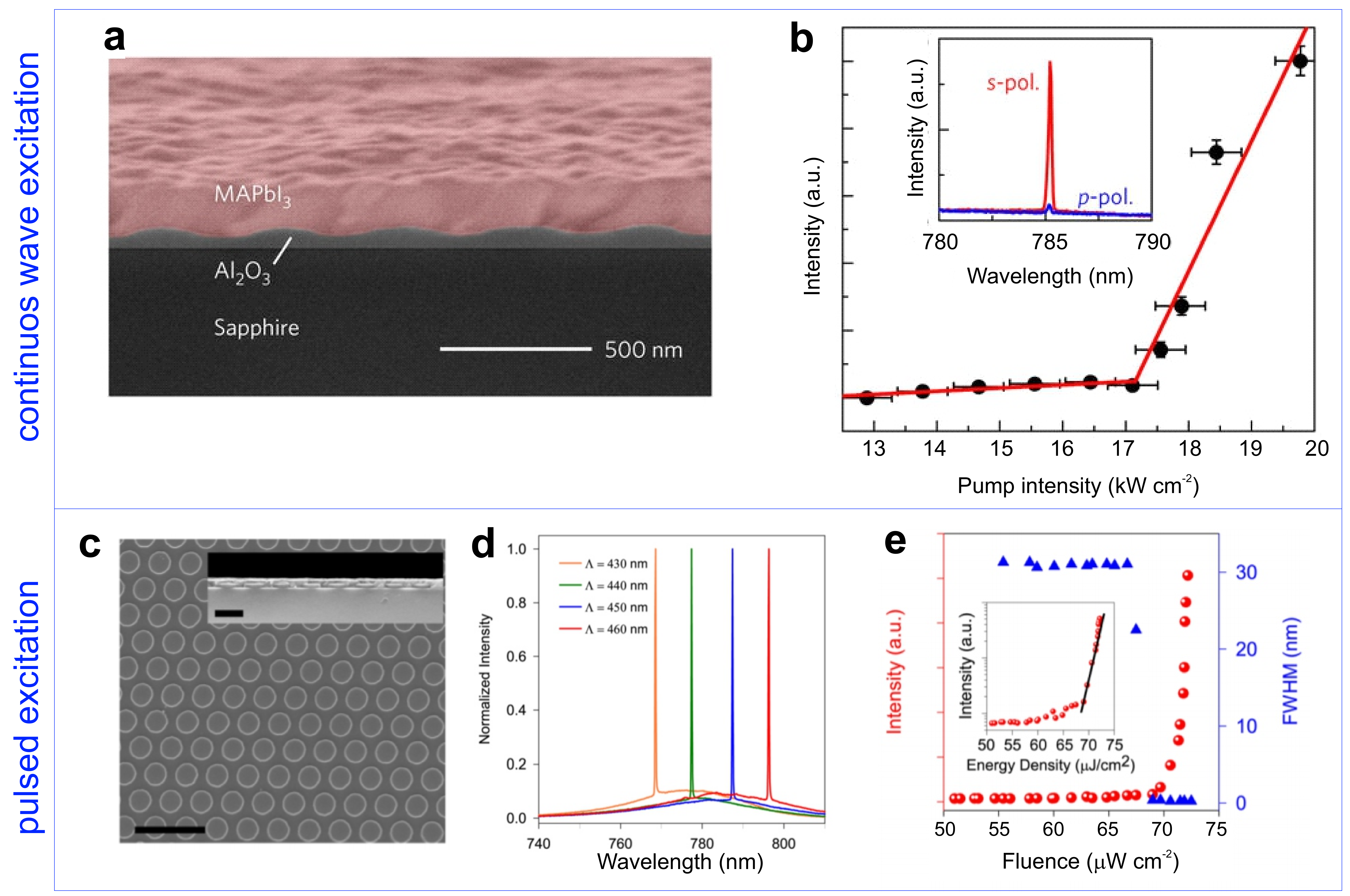}
\caption{\textbf{Lasing of perovskite photonic crystals.} (a) Colored electron image of MAPbI$_3$ perovskite-based photonic crystal, being a distributed feedback laser. (b) Experimental dependence of emission from the perovskite perforated by a sapphire grating on incident CW-laser intensity~\cite{jia2017continuous}. (c) Electron images of MAPbI$_3$ photonic crystal. The scale bars are 1~$\mu$m and 0.4~$\mu$m for main image and inset, respectively. (d) Lasing spectra from the photonic crystal with periods in range of 430--460~nm. (e) Experimental dependence of emission from the perovskite perforated by a quartz grating on incident pulsed-laser intensity~\cite{chen2016photonic}}~\label{DFB}
\end{figure*}

The strong coupling regime of light-matter interaction is very promising for creating lasers with ultra-low lasing thresholds. Polaritons undergo Bose-stimulated scattering, which surpasses spontaneous scattering at a critical density to produce the coherent condensate state, and light leaking out of the cavity from such a coherent state has been called {\it polariton lasing}~\cite{deng2010exciton}.

Lasing with high-$Q$ factors has been demonstrated in single-crystal organic-inorganic MAPbX$_3$~\cite{zhu2015lead} and all-inorganic CsPbBr$_3$ nanowires~\cite{eaton162}. Zhu {\it et al.}~\cite{zhu2015lead} demonstrated that the threshold carrier density in perovskites can be lowered significantly in single-crystal nanowires. In excitation with a \red{pulsed} laser, a \red{sharp} lasing peak was found with \red{a quality-factor of} $Q_{las}\approx$3600, which is an order of magnitude higher than that in organic semiconductor lasers. Due to a lack of grain boundaries, charge trapping in nanowires is greatly reduced and the lasing quantum yield can reach almost 100\%. 

Eaton et al.\cite{eaton162} are attributing the strong stimulated emission capability of CsPbBr$_3$ to an electron-hole plasma mechanism, which is also the leading mechanism behind ZnO and GaN lasers~\cite{eaton2016semiconductor}. The inorganic perovskite nanowires have an additional advantage of being considerably more stable in the ambient environment as compared to their organohalide counterparts~\cite{eaton162}. Typical values for nanowire-base perovskite lasers are around 0.1--10~$\mu$J/cm$^2$ under femtosecond laser pump, and 6 kW/cm$^{-1}$ which is considerably lower then those for GaN~\cite{gradevcak2005gan} and ZnO~\cite{huang2001room} nanowire lasers. For two-photon pumping of perovskite nanowires, lasing threshold is several orders of magnitude higher, e.g. up to 1~mJ/cm$^2$ for MAPbBr$_3$ under excitation by $\lambda$=800~nm femtosecond laser~\cite{gu2016two}.

Furthermore, hybride perovskite nanowires give the option to tune the emission wavelength through the entire visible spectrum~\cite{zhu2015lead, fu2016broad} and even \red{move} it to the near-IR range by introducing formamidinium (FA$^+$) instead of MA$^+$ or Cs$^+$ cations~\cite{fu2016nanowire}, as shown in Fig.~\ref{NWlas}(a). Moreover, the anion-exchange in the vapor phase allows for tuning of nanowires~\cite{he2017multi}.

Perovskite nanocavities can be fabricated not only like nanowires, but also as planar nanolasers with polygon (triangle, square, hexagon) shapes~\cite{zhang2014room}. Lasing from pyramid microcavity was studied as well~\cite{mi2018fabry}. In such microscale perovskite structures, so-called whispering gallery modes with high $Q$-factors can be excited, resulting in the effects similar to those described for nanowires, namely strong coupling and lasing~\cite{zhang2016high}, as presented in Fig.~\ref{NWlas}(b). Moreover, the nanoplates are promising for integration with advanced nanophotonic designs, as discussed below in Sec. 3.4.

\subsection{Periodic structures}

\begin{figure}
\noindent \includegraphics[width=0.95\linewidth]{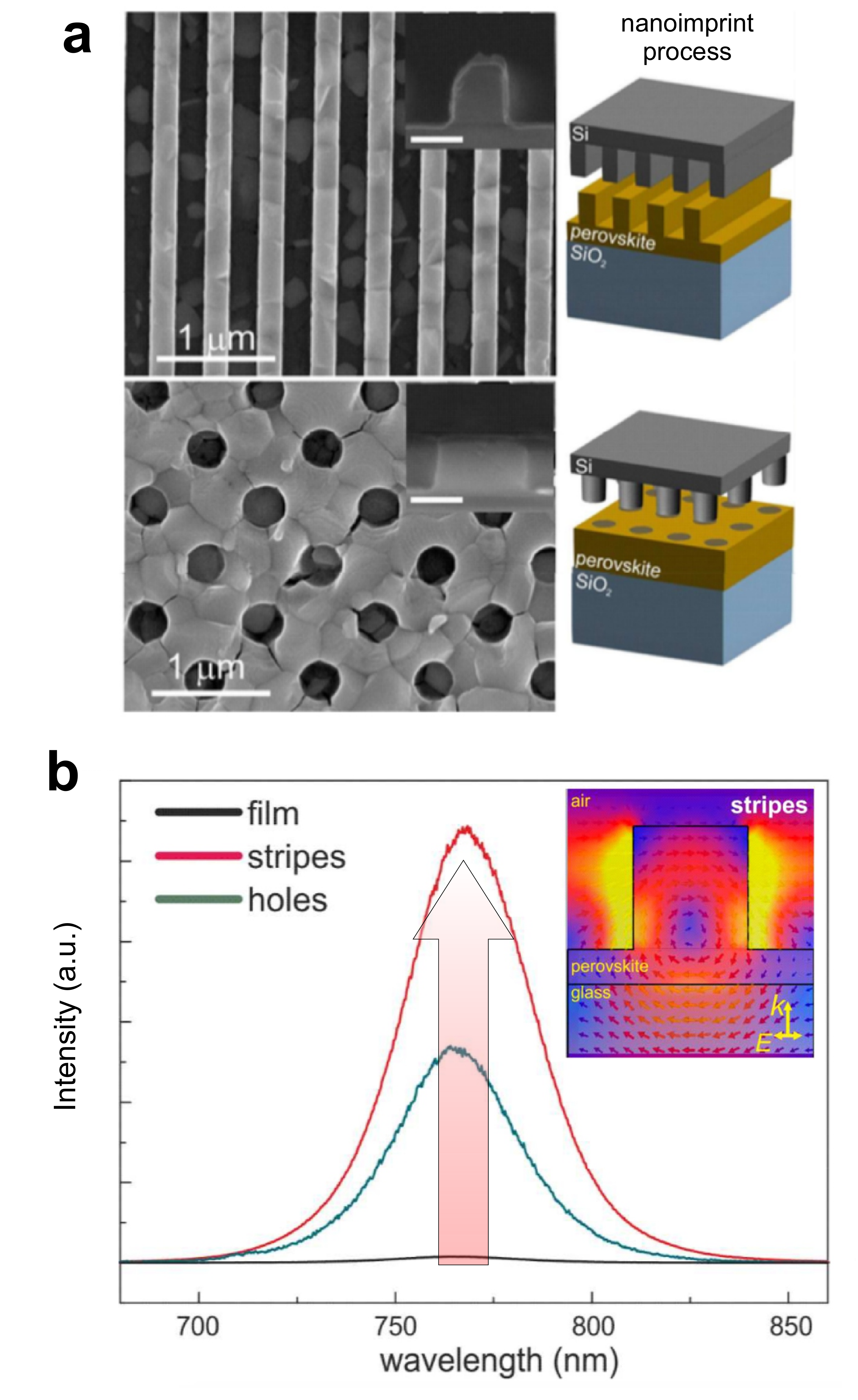}
\caption{\footnotesize{\textbf{Nanoimprinted perovskite metasurfaces for enhanced luminescence.} (a) SEM images and respective schemes of the employed nanoimprint lithographic methods. (b) Photoluminescence enhancement of different metasurfaces (nanostripes, nanoholes) over the unstructured film. An increase over 70 times can be seen between the PL of the nanostripe film over the unstructured film, see Ref.~\cite{makarov17}}}
\label{meta2}
\end{figure}

\textbf{Fabrication methods.} Direct nanoimprint of halide-perovskite films is the most common and cost-effective approach enabling high-quality reproduction of lithographycally prepared cm-scale masks~\cite{wang2016nanoimprinted, makarov2017multifold, wang2017nanoimprinted,tiguntseva2017resonant}. This method usually employs a patterned silicon mold pressing a perovskite film \red{[see Fig.\ref{meta2}(a)]} with the pressure up to 10~MPa and temperature up to 400~K for several minutes. The improvement of a crystalline structure of imprinted films was also observed~\cite{wang2016nanoimprinted}. Polymer nanopatterns, nanoimprinted~\cite{whitworth2016nanoimprinted} or etched by an electron beam~\cite{chen2016photonic} \red{were also employed as periodic substrates for further perovskites deposition}. Stability of nanoimprinted structure was improved by using triple-cation composition, e.g. Cs\textsubscript{0.05}(MA\textsubscript{0.17} FA\textsubscript{0.83})\textsubscript{0.95}Pb(I\textsubscript{0.83} Br\textsubscript{0.17})\textsubscript{3}~\cite{makarov2017multifold}.

\begin{figure}[hb!]
\noindent \includegraphics[width=0.95\linewidth]{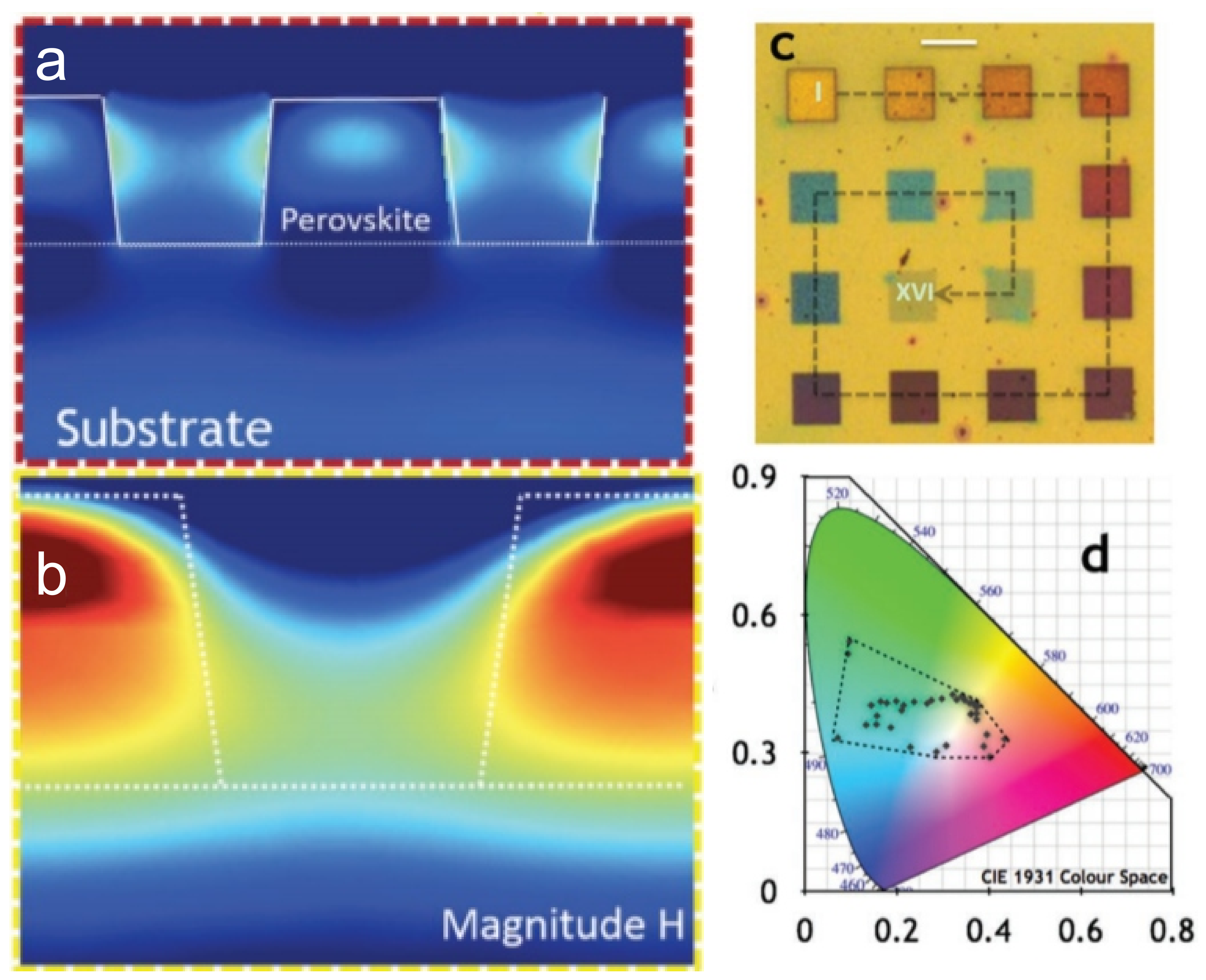}
\caption{\textbf{Perovskite metasurfaces.}  \red{Simulated (a) electric and (b) magnetic field distributions for TE polarized incident light of metasurfaces. (c) Unpolarized optical microscope image of nanograting tunable color perovskite metasurfaces 350~nm with different mill depths ranging from $\approx$200~nm to $\approx$20~nm. Scale bar indicates 20~$\mu$m. (d) CIE color palette with marked points for a selection of the nanograting metasurfaces~\cite{gholipour17}.}}
\label{meta}
\end{figure}

\begin{figure*}
\noindent \includegraphics[width=0.99\linewidth]{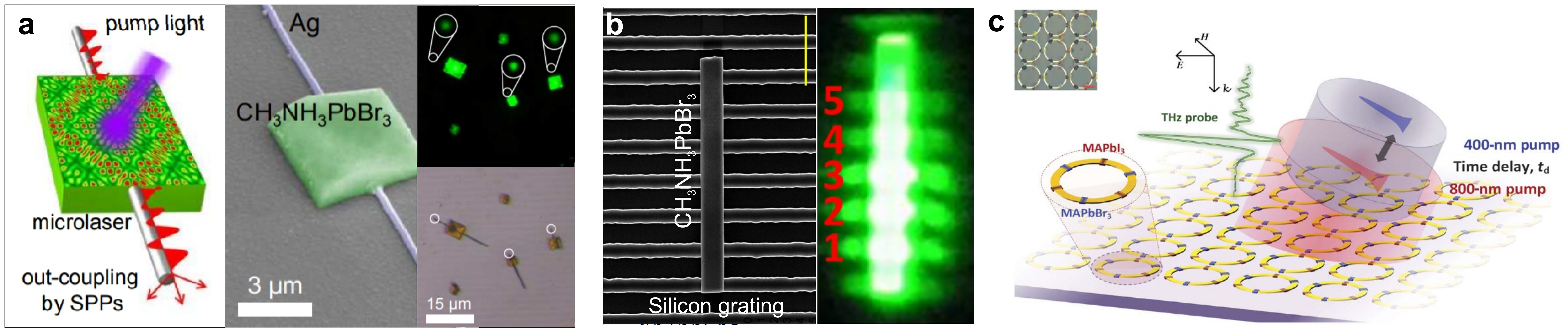}
\caption{\textbf{Integration of halide perovskites with non-perovskite nanostructures.} (a) An embedded dielectric/metal heterostructure for plasmonic output of dielectric laser. Left: schematic illustration. Central: \red{electron} image in false colors. Right: optical images in reflection and luminescent modes~\cite{li2016output}. (b) Left: \red{electron} image of the MAPbBr$_3$ microwire on silicon grating, where the scale bar is 10 $\mu$m. Right: fluorescent microscope image \cite{wang2016high}. (c) Schematic showing the time-delayed pump excitations at 800 and 400~nm, which photogenerate free carriers in MAPbI$_3$ and MAPbBr$_3$, respectively~\cite{chanana2018ultrafast}.}
\label{Int}
\end{figure*} 

{\textbf{Photonic crystals.}} Photonic crystals are periodic structures possessing band gaps for light propagating through it. One-dimensional (1D) or two-dimensional (2D) photonic crystals (the former are often called ``diffraction gratings'') are used for distributed feedback (DFB) resonators, where light propagates along a layer with higher gain~\cite{meier1999laser}. Since 1971~\cite{kogelnik1971stimulated}, the DFB lasers have become standard tools for many industrial applications, due to their low cost of fabrication, low thresholds, high quality factors, mirror-free designs, narrow linewidth, and most importantly, tunable single-mode output over a wide spectral range. Halide perovskites are suitable for DFB lasers due to their perfect characteristics as described in Sec.~2.2.

A grating-based perovskite lasers (Fig.~\ref{DFB}) were demonstrated both with CW-pump~\cite{jia2017continuous} and pulsed excitation~\cite{chen2016photonic, whitworth2016nanoimprinted}. For MAPbI$_3$, lasing threshold was reported to be 17~kW/cm$^2$ [CW excitation~\cite{jia2017continuous}, Fig.~\ref{DFB}(b)], 110~$\mu$J/cm$^2$~(ns-laser excitation~\cite{whitworth2016nanoimprinted}), 70~$\mu$J/cm$^2$ (ps-laser excitation~\cite{chen2016photonic}, Fig.~\ref{DFB}(e)), and 3--4~$\mu$J/cm$^2$(fs-laser excitation~\cite{whitworth2016nanoimprinted, pourdavoud2018distributed}). In order to lower the lasing threshold, cooling and decrease of the repetition rate from the MHz to kHz level were employed~\cite{jia2016diode}. Low-threshold amplified spontaneous emission from perovskite DFB structures were reported in Refs.~\cite{saliba2016structured, gharajeh2018continuous} .

Three-dimensional (3D) photonic crystals are usually represented by opals, which are self-assembled structures of dielectric spheres possessing a photonic band gap~\cite{blanco2000large}. Naturally occurring opals consist of submicrometer-sized air pores between the spheres. Since the size of the pores is about several hundred nanometers, they can be filled by various active materials for lasing applications~\cite{shkunov2002tunable}. Filling opals with perovskites was implemented for achieveng a strong coupling regime~\cite{sumioka01}, lasing~\cite{schnemann2017halide}, and improving solar cells characteristics~\cite{ha17}.

{\textbf{Metasurfaces.}}
Dielectric optical metasurfaces have been demonstrated for holography~\cite{wang2016grayscale}, surface coloring~\cite{proust2016all}, and many other important applications~\cite{staude2017metamaterial}. The metasurface regime corresponds to the limit when the period of a particle array ($\Lambda$) becomes smaller than the size required for diffraction, i.e. $\Lambda < \lambda_{\rm inc} /n_{1,2}$, where $\lambda_{\rm inc}$ is incident wavelength, and $n_{1,2}$ are refractive indices of the substrate or superstrate. Also, each resonant particle (the so-called ``meta-atom'') supports a Mie-type geometric resonance. The detailed discussion of the transition from te regime of photonic crystal to that of metamaterial for dielectric periodic lattices can be found in Ref.~\cite{rybin2015phase}.  

Gholipour {\it et al.}~\cite{gholipour17} demonstrated coloration of MAPbI$_3$ perovskite film surface by creating metasurfaces representing one-dimensional grating with different periods. As shown in Fig.~\ref{meta}, both electric and magnetic components of the incident light are resonantly localized in the ridges, resulting in spectrally selective enhancement of reflection and, thus, changes from blue to red colors, as shown in Figs.~\ref{meta}(c,d).
Also, such metasurfaces demonstrate up to 10-fold enhancement of photoluminescence under UV continuous wave excitation~\cite{wang17}. Similar grating [see Fig.~\ref{meta2}(a)] supporting Mie resonances in the ridges under IR femtosecond multiphoton excitation exhibited 70-fold enhancement, as shown in Fig.~\ref{meta2}(b).   Several mechanisms are responsible for the PL enhancement: (i) enhancement of linear/nonlinear absorption; (ii) the Purcell effect; (iii) improved outcopling efficiency in the detection angle.

\subsection{Other types of resonant structures}

\textbf{Integration of nanoparticles with perovskites.} One of the first integrations of resonant nanostructures with halide perovskites was demonstrated with plasmonic nanoparticles--metallic subwavelength spheroids, in order to improve the efficiency of solar cells~\cite{zhang13,saliba15,lu15,mali16,lee16,cai15,yue16}. Indeed, surface plasmons in metal nanoparticles (eg., Au, Ag) can \red{localize} incident light via optical near-field enhancement thus resulting in better absorption in a photoactive layer~\cite{jang2016plasmonic}. Detailed numerical calculations revealed the optimal parameters for increased light absorption in multilayer systems similar to a design of \red{perovskite solar-cells}~\cite{carretero16, omelyanovich2016enhancement}. Placing nanoparticles inside a layer with respect to the light source suggests that plasmonic near-field as well as scattering effects contribute to the improved conversion of light to electron-hole pairs.

The \red{adding} of plasmonic nanoparticles was found to decrease the photoluminescence intensity through the quenching effect - increase of nonradiative recombination of photo-generated carriers. Lu {\it et al.} demonstrated a nearly complete quenching of PL intensity through the inclusion of popcorn-shaped Au-Ag nanoparticles~\cite{lu15}. This effect is usually related to \red{excitation} of high-order modes in lossy metallic particles, giving a contribution to nonradiative part of Purcell factor~\cite{sun2012origin}.

Apart from \red{purely} metallic nanospheres, core-shell nanoparticles have been studied as well~\cite{zhang13, saliba15}. In such a hybrid nanoparticle,  a metallic core is surrounded by a shell made of an insulating dielectric material such as SiO$_2$ or TiO$_2$. The coating reduces the increased nonradiative recombination of photo-carriers at the metal-perovskite interface and reduces the corrosion of nanoparticles. Also, resonant silicon nanoparticles placed on a surface of a halide perovskite film have been shown to enhance both absorption and photoluminescence~\cite{tiguntseva2017resonant}, owing to resonant Mie modes and near-field enhancement, as well as low optical losses of silicon~\cite{zyuzin2018photoluminescence}.  

\textbf{Integration with other nanostructures.} 
\red{An effective} way of integration is to create nanophotonic designs based on perovskite resonant nanostructures (nanoparticles, nanowires, or nanoplates) with non-perovskite designs.

\begin{figure*}
\noindent \includegraphics[width=0.99\linewidth]{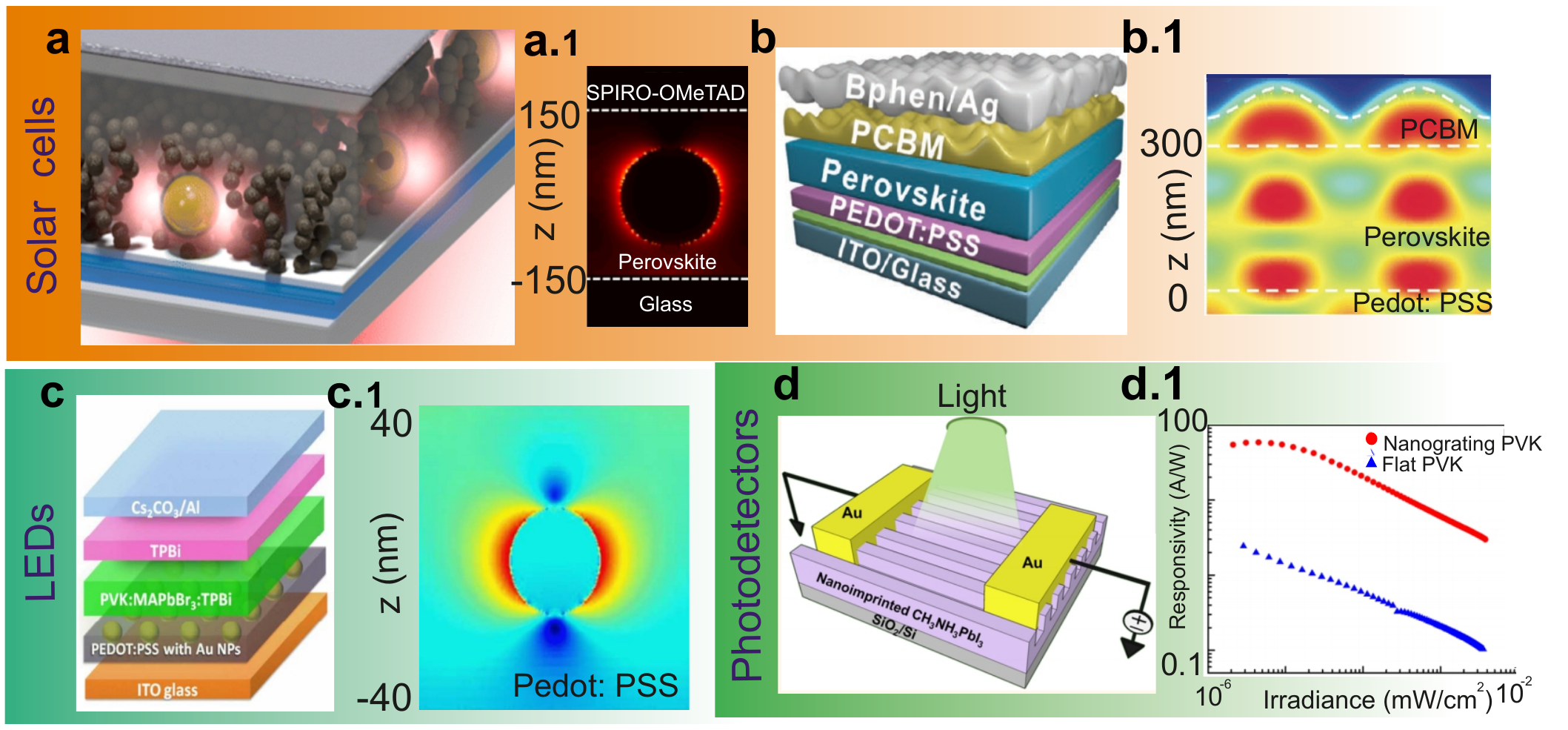}
\caption{\textbf{Integration of halide perovskites with resonant nanostructures.} (a) Gold plasmonic nanoparticles placed into m-TiO$_2$ layer of PSC. (a.1) Plasmonic nanoparticles employed for a near-field enhancement at 750 nm when being placed into perovskite. (b) Moth-eye metasurface modified PSC device and (b.1) its ability to localize the electro-magnetic field~\cite{wei2017enhanced}. (c) Implementation of gold plasmonic nanoparticles into a transport layer of perovskite-based LEDs and (c.1) FDTD stimulation of the electromagnetic field distributed around the Au nanoparticle~\cite{chen2017nearly}. (d) Nanograted perovskite metasurface of photodetector and (d.1) irradiance-dependent responsivity plot at $\lambda$ = 635 nm~\cite{wang2016nanoimprinted}}.
\label{12}
\end{figure*}

Integration of dielectric lasers with plasmonic waveguides to construct hybrid systems may help breaking the diffraction limit by localization of emitted lasing modes on the deeply subwavelength scale. Liu {\it et al.}~\cite{li2016output} demonstrated the nanoscale output of dielectric lasers via photon–plasmon coupling in rationally designed perovskite/silver heterostructures showed in Fig.~\ref{Int}(a). Here the emitted light propagates up to 10~$\mu$m along a silver nanowire, and the output coupling is modulated by controlling the resonant modes of two-dimensional perovskite microcavities, making such a design promising for ultrasmall photonic chips. Interestingly, a similar design can be based on all-perovskite platform where both nanoplate and nanowire are made of the same perovskite being arranged in a way to provide directional outcoupling of lasing modes from the nanoplate~\cite{wang2016unidirectional}. \red{In order to provide outcoupling of emission to the upper direction, nanoparticles were placed on perovskite nanoplates~\cite{wang2018dark}.}

Another approach for improved outcoupling of lasing modes from perovskite nanowires is to integrate them with a silicon grating, as shown in Fig.\ref{Int}(b), when some parts are placed on silicon ridges~\cite{wang2016high}. In this case, \red{some parts of the nanowire} do not support high-$Q$ modes due to their leakage to a high-index silicon substrate, whereas free-standing parts work as separated lasing systems. Arrays of nanolasers are also controllably created at prepatterned substrates made of silicon~\cite{liu2017organic}, polymer~\cite{he2017patterning}, sapphire~\cite{oksenberg2017surface}, and gold~\cite{huang2018formation}.

Halide perovskites can also be employed for extending functionality of plasmonic metasurfaces. This hybrid platform is prospective for ultrafast all-optical switching, where halide perovskite works as a nonlinear material (see Sec. II), and the metasurface possesses spectrally narrow optical response, which is desirable for modulation. Manjappa {\it et al.}~\cite{manjappa2017hybrid} covered a metal metasurface supporting a sharp Fano resonance by MAPbI$_3$ to achieve strong and fast (hundreds of picoseconds) all-optical modultaion of a THz signal, whereas the thickness of the structure was $\lambda$/10$^4$. Cong {\it et al.}~\cite{cong2017perovskite} applied this concept for the case of a flexible substrate.

In Fig.~\ref{Int}(c),  more complicated design of perovskite/metasurface combination is presented, where a novel fabrication technique shields already deposited perovskites from organic solvents, allowing for multiple perovskites to be patterned in close proximity~\cite{chanana2018ultrafast}. Different bandgap properties of the used hybrid perovskites (MAPbBr$_3$ and MAPbI$_3$) inserted into gaps of split-ring resonators in the metasurface also allowed for a design of THz devices that exhibit various THz responses. Here the amplitude modulation of 20\% at the timescale less than 100~ps has been achieved.

\section{Perovskite devices improved by nanostructuring}

Superior electric and optical properties of \red{halide} perovskites, along with relative simplicity of their fabrication, make them promising candidates for a novel generation of optoelectronic devices~\cite{green2014emergence, sutherland2016perovskite}. Figure~\ref{12} summarizes some major concepts of \red{the perovskite-based devices} improvement by \red{implementation} of resonant \red{nanostructuring} or integration with nanoparticles, whereas Table~\ref{devices} presents a summary of some remarkable achievements in this field. Here we overview several recently emerged nanophotonics-based low-cost approaches, which were shown to boost the optical device performance, becoming versatile tools for a broad range of perovskite compositions and device architectures.

\begin{table*}[]
\centering
\caption{Perovskite devices with external quantum efficiency (EQE) improved by resonant nanostructures: solar cells (SC), light-emiting diodes (LEDs), and photodetectors. ETL -- electron-transport layer; HTL -- hole-transport layer.}
\label{devices}
\begin{tabular}{|l|c|c|c|}
\hline
Resonant insertions                                                                                                          & Device  & Relative EQE enhancement         & Ref                                                                      \\ \hline
\begin{tabular}[c]{@{}c@{}}Au@TiO$_2$, nanospheres (80nm) in m-ETL \\ or/and in perovskite\end{tabular}                        & SC            & 44.4\%               & \cite{luo2017plasmonic}                                                \\
\begin{tabular}[c]{@{}c@{}}Au@SiO$_2$, nanospheres (14 nm) \\ between compact and m-ETLs\end{tabular}                            & SC            & 8.6\%               & \cite{aeineh2017inorganic}                                              \\
\begin{tabular}[c]{@{}c@{}}Au@SiO$_2$, nanorods (15$\times$37 nm) between\\ hole transport layer (HTL) and perovskite\end{tabular} & SC            & 30.4\%               & \begin{tabular}[c]{@{}c@{}}\cite{wu2016prominent}\end{tabular} \\
\begin{tabular}[c]{@{}c@{}}Au@TiO$_2$, nanorods (5×40) nm in ETL\end{tabular}                                                 & SC            & 7.7\%               & \cite{fan2017tailored}                                                       \\
Au (40 nm) between m-ETL and MgO                                                                                             & SC            & 34.2\%               & \cite{zhang2017efficient}                                                       \\
\begin{tabular}[c]{@{}c@{}}Au@TiO$_2$, (60 nm) on TiO$_2$ nanofibers\end{tabular}                                                 & SC            & 28.4\%               & \cite{mali2016situ}                                                      \\
\begin{tabular}[c]{@{}c@{}}Ag@TiO$_2$, (40 nm) between m-ETL and perovskite\end{tabular}                                      & SC            & 12.4\%               & \cite{xu2017surface}                                           \\
Au nanostars, (30 nm) in HTL                                                                                                  & SC     & 4.8\%, 13.2\% & \cite{ginting2017plasmonic}                                              \\
Au@SiO$_2$, (100 nm) in m-ETL                                                                                                    & SC            & 6.5\%               &  \cite{zhang13}                                                   \\
Moth-eye m-TiO$_2$ layer                                                                                                             & SC            & 11.5\%               & \cite{kang2016moth}                                           \\
grated perovskite                                                                                                            & SC            & 18\%                 & \cite{wang2018diffraction}                                           \\
grated/ moth-eye back electrode                                                                                              & SC            & 8.4\%/14\%   & \cite{wei2017enhanced}                                                   \\
Nanocone substrate                                                                                                           & SC            & 8.9\%               & \cite{tavakoli2015highly}                                                  \\
Nanocone substrate                                                                                                           & SC            & 79.4\%                 & \cite{tavakoli2016efficient}                                                   \\ 
Nanobowl TiO$_2$ layer & SC & 38.2\% & \cite{zheng16}\\
3D pyramid substrate & SC & 19.7\% & \cite{wooh13}\\
Au nanosperes (20 nm) in HTL & LED & 97\% & \cite{chen2017nearly}\\
Ag nanorods (55$\times$30 nm) between HTL and perovskite & LED & 43.3\% & \cite{zhang2017plasmonic}\\
Grated perovskite & Photodetector & 690\% on/off, 3500\% responsivity & \cite{wang2016nanoimprinted}\\
Au nanospheres (8~nm) & Photodetector & 113\% responsivity & \cite{dong2016improving}\\
Au metasurface & Photodetector & 250\% & \cite{du2018plasmonic}\\
\hline
\end{tabular}
\end{table*}

\textbf{Solar cells.}
Perovskite-based solar cells (PSC) became an explosive sensation in photovoltaics due to a sharp jump in the device efficiency from 3\% to 22\% in a few years~\cite{kojima2009organometal, lee2012efficient,yang2017iodide}, and a rapid progress in the improvement of stability~\cite{grancini2017one}. Many attempts have been undertaken to reach the theoretical Shockley-Queisser limit not only by the modification of photoactive perovskite layers but also by incorporating photoactive nanostructures, usually plasmonic nanoparticles~\cite{luo2017plasmonic, ginting2017plasmonic}, nanostructured layers~\cite{kang2016moth,wei2017enhanced} or antireflective substrates~\cite{tavakoli2015highly,tavakoli2016efficient}. Table~\ref{devices} shows the most notable results of the PSC efficiency improvement related to the integration of resonant nanostructures.

The effect of plasmonic nanoparticles on the EQE growth is associated with the near-field coupling by plasmonic nanoparticles located close to a photoactive perovskite layer~\cite{aeineh2017inorganic}, as well as improved hot-electrons transfer and charge separation~\cite{cui2016surface, mali2016situ, fan2017tailored}. We notice that plasmonic nanoparticles should be covered by an additional shell from SiO$_2$ or TiO$_2$ to avoid harmful chemical interaction with halide perovskites. Apart from plasmonic nanoparticles, it seems promising to use Mie-resonant nanostructures~\cite{kuznetsov2016optically} with hight refractive index and low losses, in order to trap light in a photoactive layer and enhance PSC characteristics. Namely, silicon nanoparticles with diameters in the range 100--200~nm increase photocurrent at 500-780~nm range and push the efficiency of MAPbI$_3$-based PSC up to 19\%~\cite{furasova2018resonant}.

Zheng {\it et al.} increased light absorption in the perovskite film through a TiO$_2$ nanobowl structure. Improved charge carrier extraction, light transmission and charge carrier transport enhanced the efficiency from 8.76\% to 12.02\%~\cite{zheng16}. Nanostructuring of transport layers and back contact help to maximize the light trapping and increase the surface area between layers~\cite{wei2017enhanced}, as shown in Figure~\ref{12}. Nanostructuring of a perovskite layer via simple nanoimprint technology improves the efficiency of MAPbI$_3$-based PSC up to 19.7\%~\cite{wang2018diffraction}. The original method of inverse opals as an enhancement for optical properties of perovskites was applied to PSCs by Ha {\it et al.} in 2017. A superlattice of densily packed TiO$_2$ spheres surrounded by MAPbI$_3$ perovskite strongly reduces the hysteresis generally seen in the I-V curve of PSCs~\cite{ha17}.

An ultra-thin nanocone-structured surface was employed as an antireflective coating instead of a flat glass to reduce the reflection of incident radiation~\cite{tavakoli2015highly}. This simple solution allowed more light to penetrate inside PSC. Apart from efficiency increasing from 12\% to 13.1\%, this antireflection coating possesses self-cleaning properties, which is significant for applications in real water-unstable perovskite devices. Replacing a glass substrate by a plastic substrate with antileflection properties~\cite{tavakoli2016efficient} makes the devices flexible and resistant to a mechanical stress. 

\textbf{Light emitting diodes.} Owing to high luminescence efficiency of halide perovskites  and high charge mobility, the perovskites are not only suitable for solar cells, but also for the reverse process suchg as the light emission~\cite{sutherland2016perovskite}. Perovskite light-emitting devices (LEDs) exhibit narrow emission spectra (with FWHM $\approx$ 10--30~nm) which are tunable through the whole extended visible range, while their spectra do not depend on the size of crystallites.

However, modifying perovskite-based LEDs by resonant nanostructures has not been developed so intensively as that for SCs. We are aware of only a few attempts to add plasmonic nanoparticles to LEDs employing CsPbBr$_3$~\cite{zhang2017plasmonic}, where the authors claim that Ag nanorods enhance electroluminescence through the plasmonic field localization and the quenching effect in the presence of lossy metal nanostructures. In another study~\cite{chen2017nearly}, it was pointed that the efficiency growth is not attributed solely to the local-field enhancement, but it is also explained by the improvement of the  electric conductivity of the transport layer provided by metal nanoparticles.

\textbf{Photodetectors.}
Similar to PSCs, there exist two main strategies for improving the performance of perovskite photodetectors: enhancing the carrier transport and light harvesting~\cite{dong2017recent}. Nanostructured perovskite photodetectors 
absorb more incident light due to the suppression if light trapping or reflectance. Wang 
{\it et al.}~\cite{wang2016nanoimprinted} implemented the nanoimprint lithography to show the 35-fold improvement of responsivity and 7-fold improvement of the on/off ratio for MAPbI$_3$ perovskite photodetectors as compared with unimprinted devices. 
Light scattering in porous CsPbBr$_3$ perovskite also helped to enhance on/off ratio and external quantum efficiency of photodetector~\cite{xue2017constructing}.

As shown by Dong {\it et al.}~\cite{dong2016improving}, incorporating plasmonic nanoparticles to the CsPbBr$_3$ photodetectors allows to improve their on/off ratio up to 10$^6$, and the photocurrent enhancement from 245.6~$\mu$A up to 831.1~$\mu$A. Lithographically fabricated arrays of gold nanoparticles covered by a MAPbI$_3$ layer form photodetectors with EQE enhanced from 15\% to 65\% in the near-IR frequency range~\cite{du2018plasmonic}.

Finally, we would like to stress specifially on the reported high values of the efficiency enhancement for perovskite-based devices as well as low cost of the employed nanophotonic approaches. Namely, colloidal nanoparticles are available on market, whereas nanoimprint technique is the most high-throughput among the known lightographic methods being fully complementary to the roll-to-roll process.

\section{Conclusion and Outlook}

We notice again that the recently emerged active study of optical properties of halide perovskites suggest them as promising materials for numerous applications in photonics. These novel opportunities appear due to a number of specific properties of perovskites such as  high refractive index  and the existence of excitons at room temperatures, combined with their low-cost fabrication, broadband tunability, high optical gain and nonlinear response, as well as simplicity of their integration with other structures, 

In this paper, we have summarized the recent advances in this field being however restricted by the study of optical effects originating from structuring of perovskites, The main results and demonstrated properties of such materials can be summarized as follows:

\begin{itemize}

\item Submicron resonant halide-perovskite particles demonstrate single-mode lasing at room temperatures in the entire visible spectral range;

\item Halide-perovskite nanowires allow realizing the strong coupling regime at room temperatures with the large Rabi splitting (exceeding 0.5~eV) and lasing with the extremely low threshold ($<$~1~$\mu$J/cm$^2$);

\item Halide-perovskite nanowires and photonic crystals support lasing under a continuous-wave optical pump at room temperatures;

\item Halide perovskite metasurfaces enable generating structural colors, and they can be fabricated by a high-throughput method of nanoimprinting;

\item Simple wet-chemistry approaches for integration of halide-perovskite nanostructures with advanced and functional designs have been developed;

\item Halide-perovskite-based photovoltaic and optoelectronic devices have been improved significantly by merging them with advanced nanophotonic designs. 

\end{itemize}

As a reason for a rapid development in this field during last few years, we notice that halide perovskites are expected to provide a novel versitile platform for photonic and optoelectronic devices and to be implemented for numerous applications. Further progress in this field is expected to address a number of challenges, including

\begin{enumerate}[label=(\roman*)]

\item Increasing stability of nanostructures made of halide perovskites, by varying their composition and coating by polymers. The use of layered perovskites (in the so-called Ruddlesden-Popper phase) with improved stability and excellent excitonic properties may provide another prospective approach.

\item Extending the emission wavelengths of halide-perovskite nanostructures both down to UV and up to IR frequency ranges, in order to use them for telecomunication and also for the integration in silicon optical circuits, as well as to use them as UV pumps for semiconductor materials.

\item Achieving ultimately small, tunable, and electrically pumped halide-perovskite nanolasers to create a new generation of highly efficient ultracompact coherent on-chip light sources for integration. 

\item Demonstrating highly efficient nonlinear light frequency conversion from UV to THz frequncy ranges based on nanostructured thin films of halide perovskites which support specifically engineered resonant modes.

\item Making advanced nanophotonic hybrid designs (such as employing colloidal nanoparticles or nanoimprinting), to boost performance of halide-perovskite devices, to make them commercially available and also complement the best device architectures.

\end{enumerate}

In addition, we expect that resonant halide-perovskite nanostructures will ultimately be employed for the use of all of their advantages in many important nanophotonics applications such as sensors, nanolasers, photodetectors, as well as all-optical chips and logic gates. 

\section*{Acknowledgements}

This work was supported by the Ministry of Education and Science of the Russian Federation (projects 14.Y26.31.0010 and 16.8939.2017/8.9), the Australian National University, and the Australian Research Council. The authors also acknowledge a partial support from the Welch Foundation (grant AT 16-17). 

\bibliography{perovskites}

\end{document}